\documentclass[manuscript]{aastex}

\slugcomment{Accepted by Publications of the Astronomical Society
of the Pacific}

\newcommand{\sca}[1]{H2RG-#1-5.0$\mu$m}
\newcommand{\scas}[1]{H2RG-S#1}
\newcommand{\nir}{$\lambda \! = \! 0.6\! - 5\!~\mu \rm m$}

\begin{document}

\title{Detectors for the James Webb Space Telescope Near-Infrared
Spectrograph I: Readout Mode, Noise Model, and Calibration
Considerations}

\author{Bernard J. Rauscher and Ori Fox\altaffilmark{1}}
\affil{NASA Goddard Space Flight Center, Greenbelt, MD 20771}
\email{Bernard.J.Rauscher@nasa.gov}

\author{Pierre Ferruit\altaffilmark{2,3}}
\affil{Universit\'{e} de Lyon, Lyon, F-69003, France}

\author{Robert~J.~Hill\altaffilmark{8},
Augustyn~Waczynski\altaffilmark{4}, Yiting~Wen\altaffilmark{6},
Wei~Xia-Serafino\altaffilmark{4}, Brent~Mott, David~Alexander,
Clifford~K.~Brambora, Rebecca~Derro, Chuck~Engler, Matthew~B.~Garrison,
Thomas~Johnson, Sridhar~S.~Manthripragada, James~M.~Marsh,
Cheryl~Marshall, Robert~J.~Martineau,
Kamdin~B.~Shakoorzadeh\altaffilmark{7}, Donna~Wilson,
Wayne~D.~Roher\altaffilmark{5}, and Miles~Smith}
\affil{NASA Goddard Space Flight Center, Greenbelt, MD 20771}

\author{Craig~Cabelli, James~Garnett, Markus~Loose,\\ Selmer~Wong-Anglin,
Majid~Zandian, and Edward~Cheng\altaffilmark{8}}
\affil{Teledyne Imaging Sensors, 5212 Verdugo Way, Camarillo, CA 93012}

\author{Timothy~Ellis, Bryan~Howe, Miriam~Jurado,\\ Ginn~Lee,
John~Nieznanski, Peter~Wallis, and James~York}
\affil{ITT Space Systems Division, 1447 St.  Paul Street, Rochester, NY,
14653}

\author{Michael~W.~Regan}
\affil{Space Telescope Science Institute, 3700 San Martin Drive,
Baltimore, MD 21218}

\author{Donald~N.B.~Hall and Klaus~W.~Hodapp}
\affil{Institute for Astronomy, 640 North A\'ohoku Place, \#209,
Hilo HI 96720}

\and

\author{Torsten~B\"{o}ker, Guido~De~Marchi, Peter~Jakobsen, and
Paolo~Strada}
\affil{ESTEC, Astrophysics Division, Postbus 299, Noordwijk, NL2200 AG,
Netherlands}

\altaffiltext{1}{Also at Department of Astronomy, University of
Virginia, P.O. Box 4000325, Charlottesville, VA 22904}
\altaffiltext{2}{Also at Universit\'{e} Lyon~1, Observatoire de Lyon, 9 avenue
Charles Andr\'{e}, Saint-Genis Laval, F-69230, France}
\altaffiltext{3}{Also at CNRS, UMR 5574, Centre de Recherche Astrophysique de
Lyon; Ecole Normale Sup\'{e}rieure de Lyon, Lyon, F-69007, France}
\altaffiltext{4}{Global Science \& Technologies, Inc., 7855 Walker
Drive, Suite 200, Greenbelt, MD 20770} 
\altaffiltext{5}{Northrop Grumman Technical
Services, 4276 Forbes Blvd., Lanham, MD 20706} 
\altaffiltext{6}{Muniz Engineering Inc., 7404 Executive Place, Suite 500,
Lanham, MD 20706}
\altaffiltext{7}{AK Aerospace Technology Corp., 12970 Brighton Dam Rd,
Clarksville, MD 21029}
\altaffiltext{8}{Conceptual Analytics LLC, 8209 Woburn Abbey Road, Glenn
Dale, MD 20769}

\begin{abstract}
We describe how the James Webb Space Telescope (JWST) Near-Infrared
Spectrograph's (NIRSpec's) detectors will be read out, and present a
model of how noise scales with the number of multiple non-destructive
reads sampling-up-the-ramp. We believe that this noise model, which is
validated using real and simulated test data, is applicable to most
astronomical near-infrared instruments. We describe some non-ideal
behaviors that have been observed in engineering grade NIRSpec
detectors, and demonstrate that they are unlikely to affect NIRSpec
sensitivity, operations, or calibration. These include a HAWAII-2RG
reset anomaly and random telegraph noise (RTN).  Using real test data,
we show that the reset anomaly is: (1) very nearly noiseless and (2) can
be easily calibrated out. Likewise, we show that large-amplitude RTN
affects only a small and fixed population of pixels.  It can therefore
be tracked using standard pixel operability maps.
\end{abstract}


\section{Introduction}

The James Webb Space Telescope (JWST) was conceived as the scientific
successor to NASA's Hubble and Spitzer space telescopes.  Of all JWST
``near-infrared'' (NIR; \nir) instruments, the Near-Infrared
Spectrograph (NIRSpec) has the most challenging detector requirements.
This paper describes how we plan to operate NIRSpec's two
2048$\times$2048 pixel, 5~micron cutoff ($\lambda_{\rm co}$=5$\mu$m),
Teledyne HAWAII-2RG (H2RG) sensor chip assemblies (SCAs)\footnote{Within
NASA, individually mounted detector arrays are typically referred to as
SCAs. In the case of NIRSpec's H2RGs, the SCA consists of HgCdTe
detectors hybridized to a readout integrated circuit and mounted on a
molybdenum base (See Figure~\ref{fig-nirspec}).} for the most sensitive
observations, and provides insights into some non-ideal behaviors that
have been observed in engineering grade NIRSpec detectors.

This paper is structured as follows.  In Section~\ref{sec-jwst}, we
provide an introduction to JWST, NIRSpec, and NIRSpec's detectors.  We
have tried to keep this discussion brief, and provide references to more
comprehensive discussions in the literature.

In Section~\ref{sec-detector-operations}, we present the NIRSpec
detector subsystem's baseline MULTIACCUM readout mode. This section
includes a detailed discussion of how total noise averages down when
multiple non-destructive reads are used sampling-up-the-ramp. MULTIACCUM
readout is quite general, and most other common readout modes, including
correlated double sampling (CDS), multiple-CDS (MCDS; also known as
Fowler-N; \citealt{fow91}), and straight sampling-up-the-ramp are
special cases of MULTIACCUM. The general NIR SCA noise model presented
in this section, see Equation~\ref{eqn-multi-noise} \&
Table~\ref{tab-model-params}, is validated using real and simulated test
data.

Where practical, our methods and conclusions are anchored by
measurement. One advantage of the NIRSpec program is that multiple test
SCAs and test facilities are available. These are described in
Section~\ref{sec-sca-summary}.

Section~\ref{sec-reset-anomaly} describes the reset anomaly as it
appears in engineering grade NIRSpec H2RGs.  The reset anomaly is fairly
well-known in the NIR detector testing community.  Here we demonstrate
using real test data that it is a nearly noise-less artifact for the
NIRSpec detectors that have been tested so far. We show that it
straightforwardly calibrates out from most science observations, and can
therefore be safely ignored by most JWST users. However, we show that
the reset anomaly can significantly bias dark current measurements if it
is not correctly accounted for. In this paper, we describe a method of
accounting for the reset anomaly in dark current measurements by fitting
a 4-parameter function to sampled-up-the-ramp pixels.

Finally, in Section~\ref{sec-rtn}, we describe what is known about
random telegraph noise (RTN) within the NIRSpec program. Using real test
data, we show that large-amplitude RTN is a property of only a small and
fixed population of pixels for the SCAs that have been
studied.\footnote{It is helpful to differentiate between large-amplitude
RTN, that would probably cause a pixel to fail to meet total noise
requirements, and the harder-to-find (but still important)
small-amplitude RTN (near the read noise floor of the SCA) that was
included in a study by \citet{bac05}. Unless otherwise indicated, we use
the acronym RTN to refer to noise that significantly exceeds the read
noise floor of the SCA. These points are discussed more fully in
Section~\ref{sec-rtn}. } Based on these data, we do not expect RTN to
significantly impact NIRSpec. While this conclusion may appear to render
studies of RTN an academic exercise, it actually mitigates that risk
that RTN could have a major impact if the affected pixels were to change
from integration to integration.

Although our discussion is focused on JWST's NIRSpec, we anticipate that
much of what we discuss will be of interest to any astronomer using
H2RGs.  The noise model is quite general, and we are aware of others
having observed both the reset anomaly and RTN. However, one caveat is
in order.  Integration and testing of the NIRSpec detector subsystem is
just beginning now.  As such, we anticipate that much remains to be
learned about NIRSpec's detectors, and that some of the specifics
presented here may change. For this reason, we have tried to focus on
general themes, rather than on the measured performance of any
particular SCA.

\section{JWST, NIRSpec, and the NIRSpec Detector Subsystem}
\label{sec-jwst}

\subsection{JWST Mission}
\label{sec-jwst-mission}

JWST is a large, cold, infrared-optimized space telescope designed to
enable fundamental breakthroughs in our understanding of the formation
and evolution of galaxies, stars, and planetary systems.  The project is
led by the United States National Aeronautics and Space Administration
(NASA), with major contributions from the European and Canadian Space
Agencies (ESA and CSA respectively).  JWST will have an approximately
6.6-m diameter aperture, be passively cooled to below T=50~K, and carry
four scientific instruments: NIRSpec, a NIR Camera (NIRCam), a NIR
Tunable Filter Imager (TFI), and a Mid-Infrared Instrument (MIRI).  All
four scientific instruments are located in the Integrated Science
Instruments Module (ISIM), which lies in the focal plane behind the
primary mirror. JWST is planned for launch early in the next decade on
an Ariane 5 rocket to a deep space orbit around the Sun-Earth Lagrange
point L2, about 1.5$\times$10$^{6}$~km from Earth.  The spacecraft will
carry enough fuel for a 10-year mission.

JWST's scientific objectives fall into four broad themes.  These are as
follows; (1) The End of the Dark Ages, First Light and Re-ionization,
(2) The Assembly of Galaxies, (3) The Birth of Stars and Protoplanetary
Systems, and (4) Planetary Systems and the Origins of Life.  Most NIR
programs will require long, staring observations, limited by the
zodiacal background at L2 in the case of NIRCam and the TFI, or by
detector noise in the case of NIRSpec.  For all of JWST's NIR
instruments, modest $\approx$100-200~kHz pixel rates will be the rule,
with total observing times per target typically $>\! 10^4$~seconds.
Teledyne H2RGs have been selected as the detectors for all three JWST
NIR instruments. For a more thorough overview of JWST, we refer the
interested reader to \citet{gar06}.

\subsection{NIRSpec}\label{sec-nirspec}

NIRSpec will be the first slit-based astronomical multi-object
spectrograph (MOS) to fly in space, and is designed to provide NIR
spectra of faint objects at spectral resolutions of R=100, R=1000 and
R=2700. The instrument's all-reflective wide-field optics, together with
its novel MEMS-based programmable micro-shutter array slit selection
device and H2RG detector arrays, combine to allow simultaneous
observations of $>$100 objects within a 3.5$\times$3.4~arcmin field of
view with unprecedented sensitivity. A selectable 3$\times$3~arcsec
Integral Field Unit (IFU) and five fixed slits are also available for
detailed spectroscopic studies of single objects. NIRSpec is presently
expected to be capable of reaching a continuum flux of 20~nJy (AB$>$28)
in R=100 mode, and a line flux of $6\times 10^{-19}~{\rm
erg~s^{-1}~cm^{-2}}$ in R=1000 mode at S/N$>$3 in 10$^4$~s.

NIRSpec is being built for the European Space Agency (ESA) by EADS
Astrium as part of ESA's contribution to the JWST mission. The NIRSpec
micro-shutter and detector arrays are provided by NASA Goddard Space
Flight Center (GSFC).

\subsubsection{NIRSpec Detector Subsystem}\label{sec-nirspec-ds}

All three NIRSpec modes (MOS, IFU and fixed slits) share the need for
large-format, high detective quantum efficiency (DQE), and ultra-low
noise detectors covering the \nir~spectral range (see Table 1). This
need is fulfilled by two $\lambda_{\rm co}\! \sim\! 5~\mu{\rm m}$ H2RG
SCAs. These SCAs, and the two Teledyne SIDECAR\footnote{SIDECAR: System
for Image Digitization, Enhancement, Control and Retrieval.} application
specific integrated circuits (ASICs) that will control them, represent
today's state-of-the-art.  This hardware is being delivered to the
European Space Agency (ESA) by the NIRSpec Detector Subsystem (DS) team
at GSFC. The DS team will deliver a fully integrated, tested, and
characterized DS to ESA for integration into NIRSpec.

The SIDECAR ASIC and NIRSpec SCA, and indeed all JWST SCAs, recently
passed a major NASA milestone by achieving Technology Readiness Level 6
(TRL-6). TRL-6 is a major milestone in the context of a NASA flight
program because it essentially marks the retirement of invention risk. 

\begin{figure} 
\begin{center}
\includegraphics*[width=4.1in]{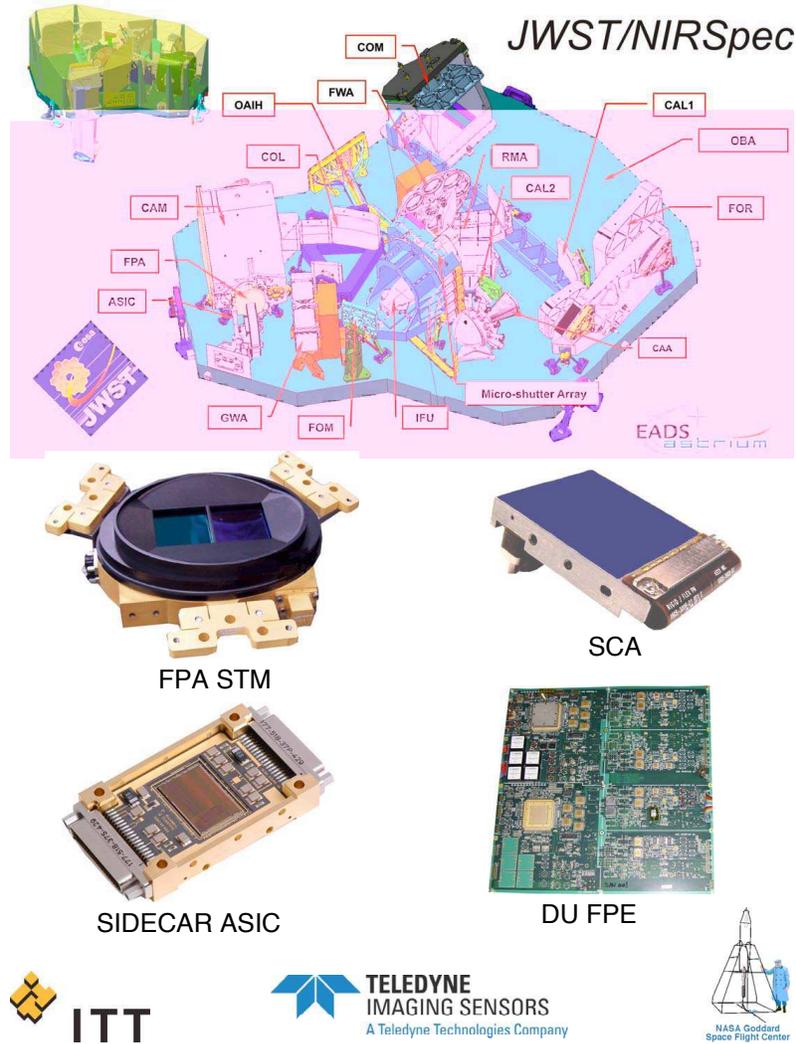}
\caption{NIRSpec is being built by EADS/Astrium for the European Space
Agency. NASA is providing the detector subsystem (DS), which is the
focus of this paper, and the micro-shutter array for target selection.
DS components include the focal plane assembly (FPA). Here we show the
structure and thermal model (STM) during test at ITT. The FPA contains
two Teledyne HAWAII-2RG sensor chip assemblies (SCAs). Other components
include two SIDECAR ASICs for FPA control and the focal plane
electronics (FPE), which control the SIDECARs. This figure shows a
development unit (DU) of the FPE undergoing test at
NASA GSFC.}\label{fig-nirspec}
\end{center} 
\end{figure}

The DS (Figure~\ref{fig-nirspec}) consists of the following components;
focal plane assembly (FPA), two SIDECAR ASICs, focal plane electronics
(FPE), thermal and electrical harnesses, and software.  The molybdenum
FPA is being built by Teledyne and their partner ITT.  The two H2RG
SCAs, which are the focus of this paper, are being built by Teledyne.

The SCA, Figure~\ref{fig-nirspec}, was designed by Teledyne and ITT.
Starting from the anti-reflection (AR) coating and going in, SCA
components include; (1) AR coating, (2) 2K$\times$2K HgCdTe pixel array,
(3) silicon readout integrated circuit (ROIC), (4) balanced composite
structure (BCS), (5) molybdenum base, (6) Rigidflex fanout circuit, and
(7) $\mu$D-37 connector.  Components 1-4 are built by Teledyne and
components 5-7 are provided by ITT.

Although NIRSpec's  DQE requirement is for \nir, the HgCdTe is actually
being grown with a somewhat longer cutoff wavelength near to
$\lambda_{\rm co}\! \sim\! 5.3~\mu{\rm m}$.  This is done to ensure
meeting the 80\% DQE requirement at $\lambda$=5~$\mu$m, and is
accomplished by varying the mole fraction of cadmium in the
Hg$_{1-x}$Cd$_{x}$Te.  In practice, proportionally less cadmium is used
to achieve longer cutoffs \citep{bri87}.

The H2RG ROIC and SIDECAR ASIC are both  reconfigurable in software. For
example, both can accommodate up to 32 video channels. For NIRSpec,
however, we plan to use only four SCA analog outputs. This is driven by
power dissipation considerations on-orbit, and by the need to minimize
system complexity. Each NIRSpec detector will return 2048$\times$2048
pixels of 16-bit data per frame. These will appear as a contiguous area
of 2040$\times$2040 photo-sensitive pixels, surrounded by a 4-pixels
wide border of non-photo-sensitive reference pixels all the way around.
Although the reference pixels do not respond to light, they have been
designed to electrically mimic regular pixels. Previous testing has
shown them to be highly effective at removing low frequency drifts like
the ``pedestal effect'' which is familiar to HST NICMOS
users \citep{are02}.

In NIRSpec, the four outputs per SCA will appear as thick,
512$\times$2048 pixels bands aligned with the dispersion direction. This
is done to minimize the possibility of calibration difficulties in
spectra that would otherwise span multiple outputs. Raw data will be
averaged in the on-board focal plane array processor (FPAP) before being
saved to the solid state recorder, and ultimately downlinked to the
ground.  The FPAP is located in the shared integrated command and data
handling system (ICDH), and is not part of the DS. Averaging is done to
conserve bandwidth for the data link to the ground.  Following
averaging, the data are still sampled-up-the-ramp, however each
up-the-ramp data point has lower noise and the ramp is more sparsely
sampled.  Detector readout will be discussed in detail in
Section~\ref{sec-detector-operations}.

Before turning to detector readout modes, it is appropriate to comment
on the performance of some prototype and engineering grade SCAs that
have been built so far.  In some cases, most notably prototype JWST SCAs
\sca{015} and \sca{006}, the parts met demanding performance
requirements including total noise per pixel, $\sigma_{\rm
total}\!<$6~$e^-$~rms per 10$^3$~seconds integration and mean dark
current, $i_{\rm dark}\!\le$0.010~$e^-$~s$^{-1}$~pixel$^{-1}$. Even with
such outstanding detectors however, getting the most out of NIRSpec will
require understanding both the ideal and non-ideal detector behaviors.

\section{Detector Readout Modes\label{sec-detector-operations}}

\begin{figure}
\begin{center}
\includegraphics*[width=5in,]{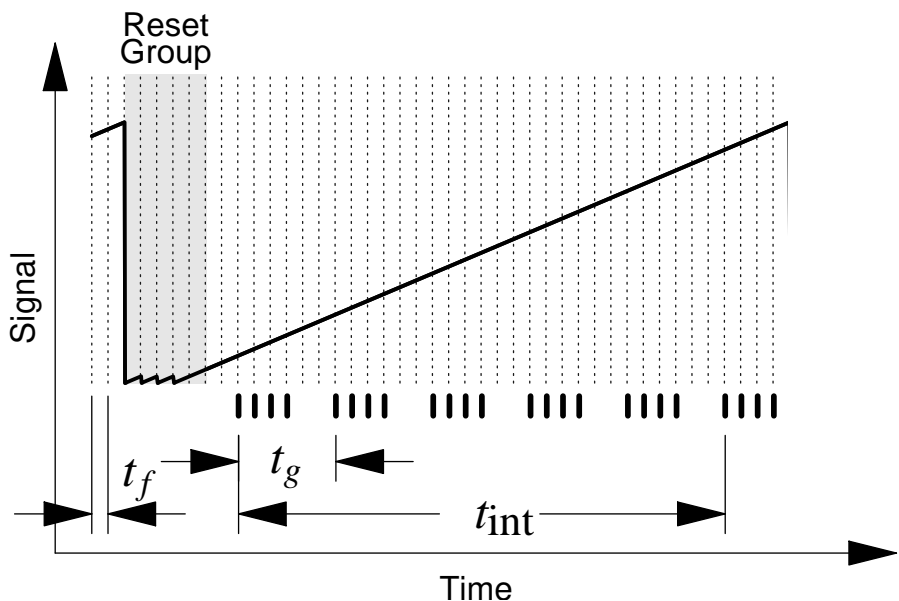} 
\caption{JWST's NIR detectors use MULTIACCUM sampling. The detector is
read out at a constant cadence of one frame every $t_f\! \approx\!
10.5~{\rm seconds}$. Although frames are clocked and digitized at a
constant cadence, to conserve data volume, not all frames are saved. In
this figure, saved frames are indicated by short, double width lines.
Likewise, to conserve downlink bandwidth, not all frames are downlinked
to the ground. Saved frames are co-added in the FPAP and averaged,
resulting in one averaged group of data being saved to the solid state
recorder every $t_g$ seconds. The resulting FITS file, consisting of a
sampled-up-the-ramp data cube with points spaced at $t_g$ intervals, is
downlinked to the ground for further processing. }\label{fig-MULTIACCUM}
\end{center}
\end{figure}

For most science observations, NIRSpec's detectors will acquire
sampled-up-the-ramp data at a constant cadence of one frame every
$\approx$10.5~s.  A frame is the unit of data that results from
sequentially clocking through and reading out a rectangular area of
pixels. Most often, this will be all of the pixels in the SCA, although
smaller sub-arrays are also possible when faster cadences are needed to
observe e.g. bright targets. Although each of JWST's NIR instruments
differs somewhat in the precise details, Figure~\ref{fig-MULTIACCUM}
shows the JWST NIR detector readout scheme.

Following in the footsteps of NICMOS, we have dubbed this readout
pattern MULTIACCUM. We frequently use the abbreviation
MULTI-$n\!\times\!m$, where $n$ is the number of equally spaced groups
sampling-up-the-ramp and $m$ is the number of averaged frames per group.
For example, in Figure~\ref{fig-MULTIACCUM}, $n$=6 and $m$=4. If a
NIRSpec user were to see a raw H2RG FITS file, it would have
dimensionality $2048\!\times\!2048\!\times\!n$. Each group, in turn, is
the result of averaging $m$ $2048\!\times\!2048$ pixel frames.

One advantage of sampled-up-the-ramp data for space platforms is that
cosmic rays can potentially be rejected with minimal data loss. Briefly
stated, we anticipate that cosmic ray hits will appear as discontinuous
steps in pixel ramps. These steps can be identified, and samples on
either side of the hit can be used to recover the slope. This has
previously been done for the HST NICMOS instrument, and we are studying
it for NIRSpec now.

In the JWST usage, the integration time, $t_{\rm int}$, is the time
between digitizing pixel [0,0] in the first frame of the first group,
and digitizing the same pixel in the first frame of the last group. The
small overhead associated with finishing the last group is not included
in the integration time.

Other important time intervals include the frame time, $t_f$, and the
group time, $t_g$.  The frame time is the time interval between reading
pixel [0,0] in one frame, and reading the same pixel in the next frame
within the same group.  The group time is the time interval between
reading pixel [0,0] in the first frame of one group, and reading the
same pixel in the first frame of the next group. For NIRSpec, the
integration time is related to the group time as follows, $t_{\rm int} =
\left(n-1\right)t_g$.

\subsection{Importance of Matching Darks/Skys}

For most astronomical NIR array detectors, it is good practice to use a
highly redundant observing strategy and matching dark/sky integrations. A
redundant observing strategy is one that samples each point on the sky
or spectrum using more than one pixel. This is usually accomplished by
building observations up from multiple, dithered integrations. The
advantage of this practice is that the non-ideal behavior of particular
pixels tends to average out, or can be identified using statistical
tools during image stacking.

Matching darks and skys are dark or sky integrations that are taken
using exactly the same readout mode as was used to obtain the science
data. For example, if the science integrations use MULTI-22$\times$4
readout, so should the darks. The same logic applies to imaging
observations of the sky. The advantage of matching calibration data is
that artifacts such as residual bias (one manifestation of the reset
anomaly, Section~\ref{sec-reset-anomaly}) subtract out.

For flight operations, one advantage of the MULTIACCUM readout pattern
is that matching darks can be easily made for all integration times if
darks are taken for the longest planned integration time. For example,
if it is known that observers will use MULTI-22$\times$4,
MULTI-6$\times$4, and MULTI-66$\times$4 integrations, a set of
MULTI-66$\times$4  darks is all that is needed for the calibration
pipeline. Darks for the shorter integration times  can be made using
only the first 22 and 6 averaged groups, respectively, from the
MULTI-66$\times$4 darks.

\subsection{Modeling MULTIACCUM Sampled Data}

In this section, we show that a general expression for the total noise
variance of an electronically shuttered instrument using MULTIACCUM
readout is, 
\begin{equation}
\sigma_{\rm total}^2 = \frac{12(n-1)}{m n(n+1)}\sigma_{\rm read}^2 +
\frac{6(n^2+1)}{5n(n+1)}(n-1)t_g f - \frac{2(2m-1)(n-1)}{m
n(n+1)}(m-1)t_f f.\label{eqn-multi-noise} 
\end{equation}
In this expression, $\sigma_{\rm total}$ is the total noise in units of
$e^-$~rms, $\sigma_{\rm read}$ is the read noise per frame in units of
$e^-$~rms, and $f$ is flux in units of $e^-{\rm~s^{-1}~pixel^{-1}}$,
where $f$ includes photonic current and dark current. The
noise model includes read noise and shot noise on integrated flux, which
is correlated across the multiple non-destructive reads
sampling-up-the-ramp. For the special case of dark integrations,
$f\!=\!i_{\rm dark}$.

Equation~\ref{eqn-multi-noise} can also be used to model CDS and MCDS
readout modes because both are special cases of MULTIACCUM.
Table~\ref{tab-model-params} summarizes the parameters to use for some
common readout schemes.  Under ultra-low photon flux and ultra-low dark
current conditions, $\sigma_{\rm CDS} \! \approx \! \sqrt{2} \sigma_{\rm
read}$.

An electronically shuttered instrument is one which does not use an
opaque shutter to block light from the detectors in normal scientific
operations.  The main exception to this rule is for taking dark
integrations.  This readout technique is in widespread use for space-based
astronomical missions, and at ground-based observatories around the
world.  In an electronically shuttered instrument, the length of an
integration is set by the readout pattern, and each pixel sees constant
flux during an integration.

JWST testing has demonstrated that dark-subtracted MULTI-$n\!\times\! m$
sampled data for a pixel, ($x$,$y$), are usually well-modeled by a 2-parameter
least-squares line fit of the form, 
\begin{equation}
s_{x,y}=a_{x,y}+b_{x,y} t,\label{eqn-linemodel}
\end{equation}
where $s_{x,y}$ is the integrating signal in units of $e^-$, $a_{x,y}$
is the y-intercept, $b_{x,y}$ is the slope, and $t$ is
time.\footnote{For example 73\% of dark subtracted pixels in engineering
grade \scas{015}, and 76\% of dark subtracted pixels in engineering
grade \scas{016} were well fitted by Equation~\ref{eqn-linemodel}. Our
criterion for ``well fitted'' is integrated chi-square probability
greater than 0.1. Of the pixels that were not well fitted, those that
we examined would have been considered inoperable because they failed
one or more operability criteria. Frequently they were obviously noisy,
with RTN being one category of noise. Although the large data sets
needed for this kind of analysis are not available for science grade
SCAs \sca{006} and \sca{015}, nothing was noted in earlier studies
suggesting that dark subtracted pixels meeting all operability are
nevertheless poorly fitted by the two-parameter model.} This point will
be elaborated on in Section~\ref{sec-reset-anomaly}. One
widely-available implementation is provided by IDL's LINFIT procedure.
In practice, however, we have found that it is much more computationally
efficient in IDL to work with full 2048$\times$2048~pixel groups of data
in parallel, and we compute the standard sums for least squares line
fitting ourselves. On our Linux and OS X computers, computing the sums
directly and in parallel is about 40$\times$ faster than calling LINFIT
sequentially for every pixel in the cube! Moreover the demands on random
access memory are greatly reduced because it is only necessary to read
in 2048$\times$2048 pixels at any one time. The expressions for the
fitted slope, $b$, and y-intercept, $a$, are as follows \citep{pre92}.
\begin{eqnarray}
 b& =& \frac{n\sum_{i=1}^n t_i s_i - \sum_{i=1}^n t_i\sum_{i=1}^n
 s_i}{n\sum_{i=1}^n t_i^2 - \left(\sum_{i=1}^n
 t_i\right)^2}\label{eqn-linefit1}\\ a& =& \frac{\sum_{i=1}^n t_i^2
 \sum_{i=1}^n s_i - \sum_{i=1}^n t_i \sum_{i=1}^n t_i s_i}{n\sum_{i=1}^n
 t_i^2 - \left(\sum_{i=1}^n t_i\right)^2}\label{eqn-linefit2}
\end{eqnarray}
In Equations~\ref{eqn-linefit1}-\ref{eqn-linefit2}, we have dropped the
($x$,$y$) subscripts for the sake of brevity. The terms $a$ and $b$ must
be computed for each pixel.

\subsection{Derivation of
Equation~\ref{eqn-multi-noise}}\label{sec-derivation}

To correctly model the noise reduction when using multiple
non-destructive reads, one must include correlated noise in the
integrating charge.  \citet{gar93} and \citet{vac04} have done this
using slightly different approaches for sampling-up-the-ramp and
MCDS readout modes.  However, the JWST readout mode is more general
than either of these. Here we extend the previous analysis to cover the
more general JWST MULTIACCUM readout mode.

In MULTIACCUM readout, the data are processed in two steps, and both are
important for correctly calculating noise correlations.  First, the data
are averaged into groups of $m$ frames in the on-board FPAP.
Subsequently, the $n$ 16-bit unsigned integer averaged groups are
downlinked to the ground for line fitting using standard 2-parameter
least-squares fitting using Equation~\ref{eqn-linefit1}.

The remainder of this section is necessarily rather mathematical.
Readers who are only interested in using Equation~\ref{eqn-multi-noise}
to model the noise of a detector system may wish to skip to
Section~\ref{sec-validation}. Here we introduce no new material, other
than that needed to arrive at Equation~\ref{eqn-multi-noise}.

Following \citet{gar93} and \citet{vac04}, the variance in the
integrated signal from continuously sampled-up-the-ramp data can be
calculated using propagation of errors as follows, 
\begin{equation} 
\sigma _{\rm total}^2 =(n-1)^2\sum _{i = 1}^n \sum _{j = 1}^n
\frac{\partial b}{\partial s_i}\frac{\partial b}{\partial s_j}C_{i, j},
\label{eqn-noise1} 
\end{equation} 
where $C_{i,j}$ is the covariance of the $j^{th}$ data point with
respect to the $i^{th}$ data point, and each $s_i$ is the average of $m$
frames.  In using Equation~\ref{eqn-noise1}, we have implicitly assumed
that each of the partial derivatives is approximately constant within
the range of variation of each $s_i$~\citep{bev69}. If this were not
true, we would have to include higher-order partial derivatives. We
therefore validate Equation~\ref{eqn-multi-noise} for the baseline
NIRSpec readout mode in Section~\ref{sec-validation}.

The covariance terms, $C_{i,j}$, are important  because the integrating
signal randomly walks away from the best fitting line as each successive
non-destructive read is acquired.  Intuitively, when frame $s_i$ is
digitized, the shot noise from frame $s_j$ is already present on the
integrating node, and we see that $C_{i,j}\!=\!s_j$ for $j\!<\!i$.
\citet{vac04} offer a simple derivation for this relation as follows.
For any two reads, $i$ and $j$, with $j\!<\!i$, the associated readout
values are $s_i$ and $s_j$, which are related by  
\begin{equation}
s_i = s_j+\Delta_{i-j}, 
\end{equation}
where $\Delta_{i-j}$ is the difference in $e^-$ between the two reads.
One can now write, 
\begin{eqnarray*} 
C_{j,i}& =&
\left<(s_j-\left<s_j\right>)(s_i-\left<s_i\right>)\right>\\ & =&
\left<s_j^2\right>-\left<s_j\right>^2+\left<s_j\Delta_{i-j}\right>-\left
<s_j\right>\left<\Delta_{i-j}\right>\\ & =&
C_{j,i}+C_{j,\Delta_{i-j}}\\ & =& \sigma_{s_j}^2\\ & =& s_j.
\end{eqnarray*}
Because integrating electrons obey Poisson statistics, we see that
$C_{i,j}\!=\!s_j$ for $j\!<\!i$.

Using Equation~\ref{eqn-linefit1}, the partial derivatives in
Equation~\ref{eqn-noise1} are found to be, 
\begin{equation}
\frac{\partial b}{\partial
s_i}=\frac{12i-6(n+1)}{n(n^2-1)}\label{eqn-partial1}. 
\end{equation}
Because $C_{i,j}=C_{j,i}$, we can rewrite Equation~\ref{eqn-noise1} as follows,
\begin{equation}
\sigma_{\rm total}^2 = \left( n-1 \right)^2 \left\{ \sum_{i=1}^n
\left(\frac{\partial b}{\partial s_i}\right)^2 C_{i,i} +  2\sum_{i=2}^n
\sum_{j=1}^{i-1} \frac{\partial b}{\partial s_i} \frac{\partial
b}{\partial s_j} C_{i,j} \right\}.\label{eqn-noise2}
\end{equation}
Using Equation~\ref{eqn-partial1}, and noting that $C_{i,i}=\sigma_i^2$
and $C_{i,j}=s_i$ where $i$ is the first of the two samples to be
acquired, Equation~\ref{eqn-noise2} can be written,
\begin{eqnarray}
\sigma_{\rm total}^2& =& \left(n-1\right)^2 \sum_{i=1}^n \left(
\frac{12i-6\left(n+1\right)}{n\left(n^2-1\right)}\right)^2
\left(\left(i-1\right)t_g f - \frac{1}{2}\left(m-1\right)t_f f
+\sigma_g^2\right) +\label{eqn-partial2}\\ && 2 \left(n-1\right)^2
\sum_{i=2}^n\sum_{j=1}^{i-1} \frac{12i-6 \left(n+1
\right)}{n\left(n^2-1\right)}
\frac{12j-6\left(n+1\right)}{n\left(n^2-1\right)}\left(j-1\right)t_g
f.\nonumber
\end{eqnarray}

In Equation~\ref{eqn-partial2}, the $\frac{1}{2}\left(m-1 \right)t_f f$
term is both important and not obvious at first glance. It comes about
because each averaged point sampling-up-the-ramp is, strictly speaking,
averaged in both the x and y-axis directions. The interval over which
shot noise is integrated therefore extends from the mid-point of one
group to the midpoint of the next. However, $\sigma_g$ already includes
the shot noise from the beginning of the group to its mid point. For
this reason, we must actually {\it subtract} the $\frac{1}{2}\left(m-1
\right)t_f f$ term in Equation~\ref{eqn-partial2} to avoid
overcompensating for this noise. Although the amount of noise accounted
for by this term is small, it shows up clearly in the Monte Carlo
simulations that were used to validate the model.

To complete the derivation, we need an expression for $\sigma_{g}$. For
the $i^{th}$ group, the FPAP performs straight 16-bit integer averaging
of the $m$ frames. 
\begin{equation}
\left<s\right>_i = \frac{1}{m}\sum _{k=1}^m s_{k,i}
\label{eqn-averaging} 
\end{equation}
For simplicity, we do not attempt to model truncation errors associated
with integer arithmetic.  As before, we use propagation of uncertainty
to write an expression for $\sigma_g$, 
\begin{equation}
\sigma_g^2 = \sum_{k=1}^m \sum_{l=1}^m \frac{\partial
\left<s\right>}{\partial s_k} \frac{\partial \left<s\right>}{\partial
s_l} C_{k,l}. \label{eqn-properr2}
\end{equation}
Because the signal within each averaged group is referenced to the first
read in that group, the reads on one group are not correlated with those
in any other. As such, all groups have the same value of $\sigma_g$.
Moreover, in this case, the partial derivatives in
Equation~\ref{eqn-properr2} are both equal to $1/m$, and using
Equation~\ref{eqn-averaging}, we can write the following. 
\begin{equation}
\sigma_{g}^2 = \frac{\sigma_{\rm read}^2}{m} +
\sum_{k=1}^m\frac{1}{m^2}\left(k-1\right)t_f f +
2\sum_{k=2}^m\sum_{l=1}^{k-1}\frac{1}{m^2}\left(l-1\right)t_f
f\label{eqn-noisepergroup} 
\end{equation}

Substituting Equation~\ref{eqn-noisepergroup} into
Equation~\ref{eqn-partial2} and simplifying, we arrive at
Equation~\ref{eqn-multi-noise}.

\subsection{Validation of
Equation~\ref{eqn-multi-noise}}\label{sec-validation}

We have validated Equation~\ref{eqn-multi-noise} using Monte Carlo
simulations, by comparing our results to others in the literature, and
by modeling real data (see Section~\ref{ra_dark}).

\subsubsection{Monte Carlo Simulations}

To validate Equation~\ref{eqn-multi-noise}, we simulated JWST NIRSpec
MULTI-22$\times$4 integrations for a range of fluxes. The simulation
parameters were as follows; $t_{\rm int}\! = \! 890.4~s$, $\sigma_{\rm
read} \! = \! 14~e^-~\textrm{rms}$, and $0.001 \le f < 64~e^-~{\rm
s^{-1}~pixel^{-1}}$. Because $f$ includes dark current, the lowest flux
simulations indicate the ultimate noise floor of the system, while
higher flux pixels indicate what might be seen when observing bright
stars.

2048$\times$2048~pixel data cubes were simulated by incrementally adding
integrated flux one frame at a time. The integrated flux during any one
frame time was distributed according to the Poisson distribution. Once
all  flux had been accumulated, normally distributed read noise was
added to all pixels in all frames. Following plans for JWST operation,
the data were then rebinned into $n$ groups of $m$ averaged frames.
Finally, Equation~\ref{eqn-linefit1} was used to compute pixel slopes,
these were converted into integrated signal by multiplying by the
integration time, and finally the standard deviation of each
2-dimensional 2048$\times$2048~pixel image was calculated.

\begin{figure} 
\begin{center}
\includegraphics*[angle=0.0,width=4.7in]{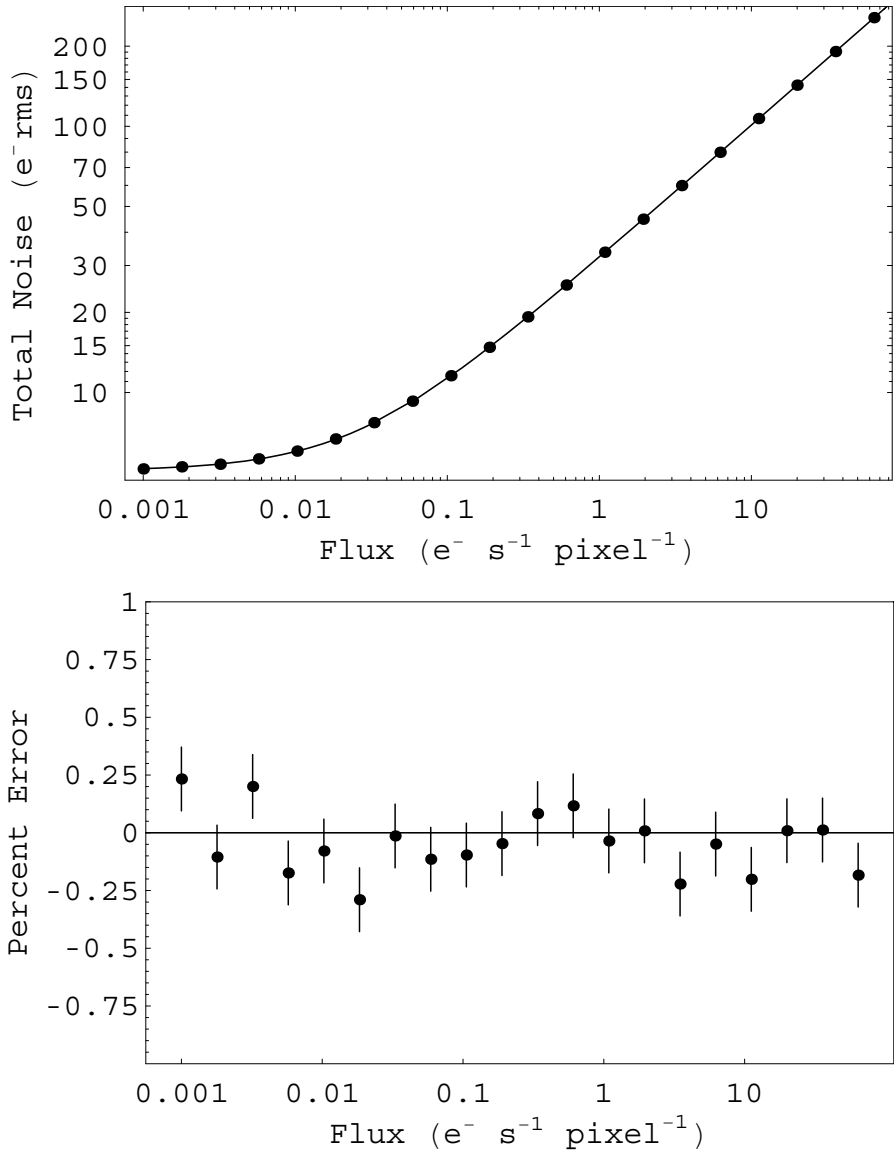} 
\caption{Equation~\ref{eqn-multi-noise} was validated using Monte Carlo
simulation of NIRSpec's MULTI-22$\times$4 readout mode. The
integration time was $t_{\rm int}=890.4~s$, the read noise was
$\sigma_{\rm read}=14~e^-$, and dark current is included in the flux, $f$. The
top panel shows total noise computed using
Equation~\ref{eqn-multi-noise} (solid line) and data points from 20
Monte Carlo simulations using approximately $10^6$ pixels per
simulation. The bottom pane shows the percent error computed under the
assumption that the Monte Carlo points represent truth.}
\label{fig-monte-carlo} 
\end{center}
\end{figure}

The results, see Figure~\ref{fig-monte-carlo}, are in excellent
agreement with Equation~\ref{eqn-multi-noise}, with all deviations
within the statistical uncertainty of the Monte Carlo simulation.

\subsubsection{Comparison to other Authors}\label{sec-litcmp}

It is helpful to consider a few limiting cases for comparison to
previous literature results.  For the case $m\!=\!1$, straight
sampling-up-the-ramp, both \citet{gar93} and \citet{vac04} contain
results that can be compared to our Equation~\ref{eqn-multi-noise}. In
particular \citeauthor{vac04}'s Equation~53 is in complete agreement with our
result.

In a similar manner, \citet{gar93} computed the total noise in read
noise dominated and shot noise dominated regimes for continuous
sampling-up-the-ramp. For read noise dominated observations, the noise
computed using Equation~\ref{eqn-multi-noise} is, 
\begin{equation}
\lim_{f \rightarrow 0} \sigma_{\rm total}^2 = \frac{12 \left(n-1
\right)}{n \left(n+1 \right)}\sigma_{\rm read}^2 , \textrm{where
}m=1\label{eqn-garcmp1}.
\end{equation}
For the shot noise dominated regime Equation~\ref{eqn-multi-noise}
becomes, 
\begin{equation}
\lim_{\sigma_{\rm read} \rightarrow 0} \sigma_{\rm total}^2 = \frac{6 \left(n^2+1
\right)}{5n \left(n+1 \right)} \left(n-1 \right)t_g f, \textrm{where
}m=1.\label{eqn-garcmp2} 
\end{equation}
Equations~\ref{eqn-garcmp1} and \ref{eqn-garcmp2} should compare to
\citeauthor{gar93}'s Equations~19 and 23 multiplied by $T_{\rm int}^2$.
However, they do not, and the difference lies in differing definitions
of the integration time. In \citet{gar93}, the integration time, $T_{\rm
int}$, is defined as the entire integration time on the detector node,
beginning when the reset switch is opened and ending when the final
signal level is sample. For most astronomical instruments, this is not
correct, and the integration time should be defined as shown in
Figure~\ref{fig-MULTIACCUM}.

Expressing $t_{\rm int}$, the correct integration time in terms of the
integration time in Garnett and Forrest's notation, $T_{\rm int}$, we
find,
\begin{equation}
t_{\rm int} = T_{\rm int} - \delta_t = T_{\rm int}
\frac{n-1}{n}, 
\end{equation} 
where $\delta_t$ is the time between successive pedestal or signal
samples. With this correction to \citeauthor{gar93}'s Equations~19 and
23, our Equations~\ref{eqn-garcmp1}-\ref{eqn-garcmp2} are in complete
agreement with theirs. For completeness, we note that a similar error
exists in \citeauthor{gar93}'s results for Fowler sampling. A correction
of the form,
\begin{equation} t_{\rm
int} = T_{\rm int} - \delta_t = T_{\rm int} \left(1-\frac{1}{2 n}
\right), 
\end{equation} 
should be made to their results for Fowler sampling.

\subsection{Effect of Neglecting Covariance Terms}

If covariance terms in Equation~\ref{eqn-noise1} are neglected,
Equation~\ref{eqn-multi-noise} simplifies as follows,
\begin{equation}
\tilde{\sigma}_{\rm total}^2 = \frac{12(n-1)}{m n(n+1)}\sigma_{\rm read}^2 +
(n-1)t_g f,\label{eqn-multi-noise2}
\end{equation}
where we have introduced the new symbol, $\tilde{\sigma}_{\rm total}$,
to unambiguously represent the approximate noise. The first term
represents read noise being averaged down, and the second term accounts
for shot noise on integrated flux under the {\em incorrect} assumption
that noise in the multiple non-destructive reads is uncorrelated. 

In the following, we consider two limiting cases: (1) the read noise
dominated regime and (2) the shot noise dominated regime. In both cases,
we compare the total noise per pixel computed using
Equation~\ref{eqn-multi-noise} to that computed using the approximate
relation, Equation~\ref{eqn-multi-noise2}.

\subsubsection{Read Noise Dominated Regime}

We first consider the read noise dominated regime. This applies, for
example, when measuring the total noise of an SCA having little or no
dark current under ultra-low photon flux conditions. JWST SCA \sca{015}
was a good example, having dark current $\le\!0.006~e^-~{\rm
s^{-1}~pixel^{-1}}$ when tested at the University of Hawaii and at the
Space Telescope Science Institute/Johns Hopkins University
\citep{rau04,fig04}. We adopt as our metric the ratio $\xi = \sigma_{\rm
total} / \tilde{\sigma}_{\rm total}$. For the read noise dominated case,
this simplifies to
\begin{equation}
\xi = \lim_{f \rightarrow 0} \frac{\sigma_{\rm
total}}{\tilde{\sigma}_{\rm total}} = 1,
\end{equation}
and we see that neglecting the covariance terms does not cause
significant errors in this case.

\subsubsection{Shot Noise Dominated Regime}

In the shot noise dominated regime, the situation is very different.
Making the simplifying assumption $m\!=\!1$, we compute $\xi$ for
straight sampling-up-the-ramp.
\begin{equation}
\xi = \lim_{\sigma_{\rm read} \rightarrow 0} \frac{\sigma_{\rm
total}}{\tilde{\sigma}_{\rm total}} = 1.095\sqrt{\frac{n^2+1}{n \left(
n+1 \right)}}, {\rm with}~m=1.\label{eqn-xi2}
\end{equation}
From Equation~\ref{eqn-xi2}, we see that for large $n$ and in the shot
noise dominated regime, Equation~\ref{eqn-multi-noise2} under-estimates
the total noise by 9.5\%. As a cross check, we note that this result is
consistent with \citeauthor{gar93}'s Equation~24. Because of this
significant error using Equation~\ref{eqn-multi-noise2}, it is
particularly important to use Equation~\ref{eqn-multi-noise} for
modeling sampled-up-the-ramp data when shot noise is important. For
completeness, in the baseline NIRSpec MULTI-22$\times$4 readout mode and
in the shot noise dominated regime, $\xi\!=\!1.071$ and we see that
Equation~\ref{eqn-multi-noise2} under-estimates the noise by 7.1\%.
Equation~\ref{eqn-multi-noise} should clearly be used in this case.

\section{Summary of Available SCAs and Test Facilities}
\label{sec-sca-summary}

The JWST Project began working with Teledyne\footnote{Teledyne Imaging
Sensors was formerly known as Rockwell Scientific. To avoid confusion,
we will exclusively use the name Teledyne when referring to the company
that is making JWST's NIR SCAs.} on the H2RG SCA for space-astronomy in
1998. Two pathfinder SCAs were produced during the development program.
These were the 1024$\times$1024 pixel HAWAII-1R, the first Teledyne SCA
to incorporate reference pixels in the imaging area, and the
1024$\times$1024 pixel HAWAII-1RG, which added a programable guide
window. Although the guide window will be used to some extent by all
JWST NIR instruments, it will be most heavily used by the TFI.

Beginning in late 2002, the first science grade H2RGs began to be
produced. For purposes of this article, a science grade SCA is one that
has excellent performance, but is nonetheless non-flight grade. Reasons
why a part might be science grade, instead of flight grade, include
differences in packaging and changes in the fabrication process.
Table~\ref{tab-sca-summary} summarizes the properties of all of the SCAs
that we discuss in this article. The two science grade parts had serial
numbers \sca{006} and \sca{015}. \sca{006} was a fully substrate-removed
part whereas the substrate-on \sca{015} was only thinned. Although these
two detectors were tested extensively at Teledyne, the University of
Hawaii, and at STScI/JHU, these early tests did not include the
extensive sets of darks that are needed for the statistical analysis
presented in Sections~\ref{sec-reset-anomaly} and \ref{sec-rtn}.

Beginning in 2006, the NIRSpec DS team at GSFC began to receive
engineering grade NIRSpec SCAs. Because the packaging was somewhat
different to that used earlier, Teledyne hybridized the lowest graded
HgCdTe layers first. These lower grade layers have yielded engineering
grade detectors with dark current and total noise exceeding NIRSpec
requirements. However, these engineering grade detectors were also the
first to be used in a fully flight representative MULTI-22$\times$4
readout mode, and with 50 ramps used for each dark current and total
noise test. Where possible, we have cross-checked our conclusions based
on the large data sets by comparison to available data from the earlier
science grade SCAs. For this reason, although the specific performance
parameters of these engineering grade SCAs are not fully flight
representative vis-$\grave{\rm a}$-vis dark current and total noise, we
believe that the general conclusions regarding the reset anomaly and RTN
are valid. As new and better SCAs arrive, we plan to continue testing
these parameters and others to enable the best possible ranking for
flight selection.

\subsection{Test Facilities}

Throughout this article, we refer freely to data acquired in the
following test laboratories.
\begin{enumerate}
\item{NASA GSFC Detector Characterization Laboratory}
\item{Teledyne Imaging Sensors Test Facility}
\item{University of Hawaii Test Facility}
\item{Operations Detector Laboratory at STScI/JHU}
\end{enumerate}
In this section, we briefly describe the equipment used in each of these
laboratories. We begin, however, with a short discussion of conversion
gain, which is used to convert from instrumental analog to digital
converter units (ADUs) to electrons. This important parameter is
measured by all NIRSpec test laboratories.

\subsubsection{Conversion Gain\label{sec:g}}

In recent years, it has become increasingly clear that inter-pixel
capacitance (IPC) can significantly affect the conversion gain of hybrid
detector arrays like the H2RG \citep{moo04,moo06,bro06}. For this paper,
which is based on archival data, the photon transfer method was used to
measure conversion gain in all laboratories \citep{jan87}, and no
correction for IPC was made. Based on our own preliminary IPC
measurements, and \citet{bro06}'s results for a $\lambda_{\rm
co}=1.7~\mu{\rm m}$ SCA, we believe that this results in systematic
over-estimation of the conversion gain (in units of $e^-{\rm~ADU^{-1}}$)
by about 10\%-20\% for the measurements that are reported in this
article. In other words, the measurements that we report here probably
over-estimate the noise, dark current, and DQE by 10\%-20\%. 

For the NIRSpec, we plan to measure IPC by using the H2RG SCA's
individual pixel reset capability to directly program pixels to
different voltages than their neighbors. We believe that this will allow
us to directly measure the crosstalk, and thereby the IPC. This
capability is being implemented now, and we plan to begin phasing it
into NIRSpec testing starting in late 2007.

\subsubsection{NASA GSFC Detector Characterization Laboratory}
\label{dcl}

The NASA GSFC Detector Characterization Laboratory (DCL) is a facility
for the design, integration, test, and characterization of detector
systems. Major projects include testing detectors for the NIRSpec DS and
the Hubble Space Telescope Wide Field Camera 3. The DCL facility that
will be used for testing the integrated NIRSpec DS consists of a Class
100 (ISO Class 5) cleanroom and a nearby test control room. The
cleanroom houses the test dewar (containing the FPA and SIDECAR ASICs),
the room temperature FPE, laboratory array controllers, dewar
temperature controllers, optical sources, dewar control, monitoring, and
interface electronics, and other support hardware.  The control room
houses test control and analysis computers, including a Science
Instrument Development Unit (SIDU) and a Science Instrument Integrated
Test Set (SITS) that communicate with and command the DS. The SIDU and
SITS mimic the functionality of the ICDH to facilitate ground-based
testing.

The dewar is a custom designed and built cryocooled system from Janis
Research Company, Inc. (Model: Pulse Tube Dewar, Serial Number 8862-B). 
The cooling is provided by a two-stage Cryomech, Inc. Model PT407 pulse
tube cryorefrigerator.  The dewar is designed to accommodate a NIRSpec
FPA containing two Teledyne H2RG SCAs, two Teledyne SIDECAR ASICs, and
two NIRSpec flight-design ASIC-to-SCA cables.  The temperatures of  the
mounting fixtures to which the FPA and ASICs mount are independently
controlled by heaters and thermometers.  The FPA and ASIC mounting plate
temperature control, as well as the dewar housekeeping temperature
control and monitoring, is provided by LakeShore Cryotronics, Inc.
temperature controllers (one model 331 and two model 340s).

Non-flight-design cables connect the ASICs and the FPA thermal control
circuits to hermetic connectors on the dewarÕs vacuum shell.  External
cables connect the ASICs and FPA thermal control circuits to the FPE. 
The FPE communicates to the SIDU or the SITS in the control room via
Spacewire cables.

For the initial SCA-level tests that are discussed in this paper and
diagnostics, another cable is available inside the dewar to bypass the
ASIC and ASIC-to-SCA cable, and connect directly to either SCA to allow
operating that SCA with laboratory electronics.  The laboratory
electronics are Generation III controllers from Astronomical Research
Cameras, Inc. Within the NIR detector testing community, these are
colloquially referred to as ``Gen-III Leach Controllers.'' For this
paper, a video gain of about 40$\times$ was used, resulting in a median
conversion gain, $g\approx~0.9~e^- {\rm ADU^{-1}}$. For SCAs H2RG-S015
and H2RG-S016, the photon transfer method was used to measure the
conversion gain of each part. For these parts, the measured median
conversion gains were $g=0.89$ and $0.93~e^-~{\rm ADU^{-1}}$
respectively. For the testing reported here, the DCL clocked SCAs at 100
kHz per pixel, and the video bandwidth was limited to about 160 kHz
using RC filters on the inputs.

\subsubsection{Teledyne Imaging Sensors Test Facility}\label{tis-lab}

Teledyne Imaging Sensors has developed an infrared detector testing
facility to support production testing and flight detector selection for
the JWST program.Ê This focus puts emphasis on test throughput,
repeatability, and flight documentation.Ê The importance of test
throughput is easy to see by looking at the JWST test requirements.Ê
ÊThe three instruments using HgCdTe detectors on JWST will be producing
approximately 180 SCAs for testing.Ê Of these, approximately 20 will be
selected as flight-quality.Ê ÊThe time period for testing and
flight-device selection is only about 1 year.Ê ÊRepeatability of
measurements requires a rigorous program of calibration and
verification, and includes cross-checking with external laboratories
using both reference diode and SCA standards.ÊÊTo eliminate the
possibility of operator variability, a highly automated system of
acquisition, analysis, and reporting has been implemented.Ê Lastly,
since the SCAs are to be selected for space flight use, significant
effort is spent on configuration management, environmental controls,
contamination monitoring and control, and documentation.

Three cryostats perform all the testing for JWST.Ê Each of these
cryostats can accommodate up to four H2RG sensors in one cooldown.Ê In
practice, one of the SCA positions is frequently allocated to a
``control'' SCA or reference diode to verify test consistency.Ê All of
these cryostats are custom designs, and operated with custom electronics
and software.Ê Their internal design is such that light-tight labyrinths
are included at all mechanical interfaces, consistent with the need for
low-background performance at $\lambda\! =\! 5 \mu {\rm m}$ $\left( f\!
<\! 0.01~e^- {\rm ~s^{-1} ~pixel^{-1}} \right)$.Ê Cooling is provided by
CTI mechanical cryocoolers, with the compressors located in the
mezzanine above the laboratory.Ê Each cryostat has three separately
controlled temperature zones that are cooled from a two-stage cold
head.Ê These zones provide for a $\sim 30$~K inner radiation shield, the
77~K outer radiation shield, and the SCA temperature (typically 37~K).

For low noise testing, the custom readout electronics are operated at a
100~kHz per pixel readout rate and the video bandwidth is limited to
about 160~kHz. The video gain of 40$\times$ and 5~Volt analog-to-digital
converters combine to yield a typical conversion gain of $\sim\!
0.477~e^-~{\rm ADU^{-1}}$.

The cryostats have two basic configurations.Ê The ``Duomo''
configuration has the SCAs viewing a short, squat diffuse-gold dome that
is illuminated by internal LEDs.Ê For each wavelength, there are 4 LEDs
illuminating the dome at 90$^\circ$ azimuthal spacing.Ê There is enough
room around the dome to place LEDs for 7 distinct wavelengths.Ê Because
the entire SCA and dome configuration can be cooled to the 37~K
operating temperature, this configuration provides the ultimate in dark
current capability.Ê Because the LEDs are illuminating the SCAs almost
directly, there is very little attenuation of the flux.Ê Two of the
three cryostats are typically used in this configuration, which is
capable of demonstrating all flight requirements except for the most
stringent DQE measurements.Ê These are limited by the illumination
uniformity at the SCAs from this physically compact arrangement
(approximately 10 to 15\% variability from center to corner) and also
the calibration uncertainty of the measurement (typically $\sim$
5\%).

The second configuration is ``Il Campanile.''Ê This uses the same
configuration of the cryostat as Il Duomo for housing and cooling the
SCAs, except that the illumination now comes from a small aperture
$\sim$500~mm away from the SCAs.Ê The aperture is fed by an integrating
sphere, which in turn is fed by LEDs.Ê The size of the aperture is
adjusted to provide the desired intensity of illumination.Ê There are
again 7 distinct LEDs that can be commanded to illuminate the
integrating sphere.Ê Carefully designed baffles and light traps
eliminate stray light.Ê The Il Campanile configuration requires a
second, single-stage, cold head for cooling the illumination components
to $\sim$77~K.

In normal usage, Il Duomo configurations are used to screen incoming
detectors for key performance parameters.Ê The acceptance thresholds
(especially for DQE)Ê are set generously in order to avoid discarding
potentially acceptable devices.Ê The exact level depends on program
requirements, taking into consideration the typical measurement accuracy
of the system.Ê After this initial screening, devices that are
potentially flight-grade go through a two week period of
characterization, at the end of which all performance parameters are
reported.Ê For programs requiring DQE measurements better than the
$\sim$ 15\% level, the best devices are placed in Il Campanile for DQE
characterization that can take up to one week.Ê Typical accuracies are
wavelength-dependent, but are on the order of 5 to 10\%.

For short-wave $\left( \lambda_{\rm co}\! =\! 2.5~\mu{\rm m} \right)$
devices, both configurations are sufficiently dark to confirm
performance to JWST levels.Ê However, because the Il Campanile has a
large physical extent, cooling the baffles and supporting structure to
less than $\sim$70~K is impractical.Ê Consequently, for the mid-wave
$\left( \lambda_{\rm co}\! =\! 5~\mu{\rm m} \right)$ devices, the Il
Campanile configuration will be too warm to reach flight performance
levels, but is more than adequate for DQE measurements.

While the main application for these cryostats is JWST testing, they
have been successfully used to support other astronomy (low-background)
programs, as well as for internal process-development testing.Ê The
cryostat design is sufficiently modular to support the differences in
mechanical mounting, heat straps, connector pinouts, etc.,Ê that could
be required for testing many kinds of devices. ÊThis flexibility also
drives the need for strict configuration management during production
testing, as well as a certification program for the test stations after
configuration changes.

\subsubsection{University of Hawaii Test Facility}

The University of Hawaii laboratory was the first test facility to
convincingly demonstrate the ultra-low dark current and noise properties
of Teledyne $\lambda_{\rm co}\! =\! 5~\mu{\rm m}$ HgCdTe for JWST. These
early tests were done using a cryocooled dewar, LakeShore temperature
controllers, and a modified Leach controller. Although the University of
Hawaii is now testing using SIDECAR ASICs in lieu of Leach controllers,
this paper is based on archival data that were taken before the SIDECAR
became available. When testing with the Leach controller, the University
of Hawaii typically reads out SCAs at a 100~kHz per pixel rate. The
video bandwidth is limited to about 160~kHz, and when operated at
40$\times$ video gain, the conversion gain is about $1~e^-~{\rm
ADU^{-1}}$.

For more information about the University of Hawaii test facility, the
interested reader is referred to the following publications
\citep{hal00,hal04,hal06}.

\subsubsection{Operations Detector Laboratory at STScI/JHU}

The Operations Detector Lab (ODL) is a joint Space Telescope Science
Institute/Johns Hopkins University facility. The primary goal of the
ODL is to be able to test flight-like JWST and HST detectors to
determine the best way to operate the detectors in flight. This is a
different focus that the other JWST labs in that the lab does not try to
verify requirements, but instead has the goal to optimize the total
science output from the instruments.

Currently, the lab has one IR Labs dewar that uses a CTI model 1050
cryo-cooler to cool both the SCA and internal optics to their
operational temperatures (nominally 37 and 60~K respectively). A
LakeShore model 340 temperature controller is used to stabilize the
temperature of the SCA to within $<$1~mK per 1000 seconds. A variety
of optical configurations are available to either allow direct imaging
with a Offner relay, a pinhole camera, or a cryogenic integrating
sphere. The detector is housed in a light-tight enclosure where the
upper limit on the light leak is 1 photon per 1000 seconds.

The readout electronics use a Generation II controller from Astronomical
Research Cameras Inc. Pixels are read out at a 100~kHz per output rate,
and the video bandwidth is limited to about 160~kHz using RC filters.
The baseline video gain is 40$\times$ and the measured conversion gain,
$g\! \approx\! 1~ e^- ~{\rm ADU^{-1}}$.

For more information on the ODL's test setup, the interested reader is
referred to \citet{fig03}.

\section{Reset Anomaly}
\label{sec-reset-anomaly}

\begin{figure}
\begin{center} 
\includegraphics*[angle=0,width=5in]{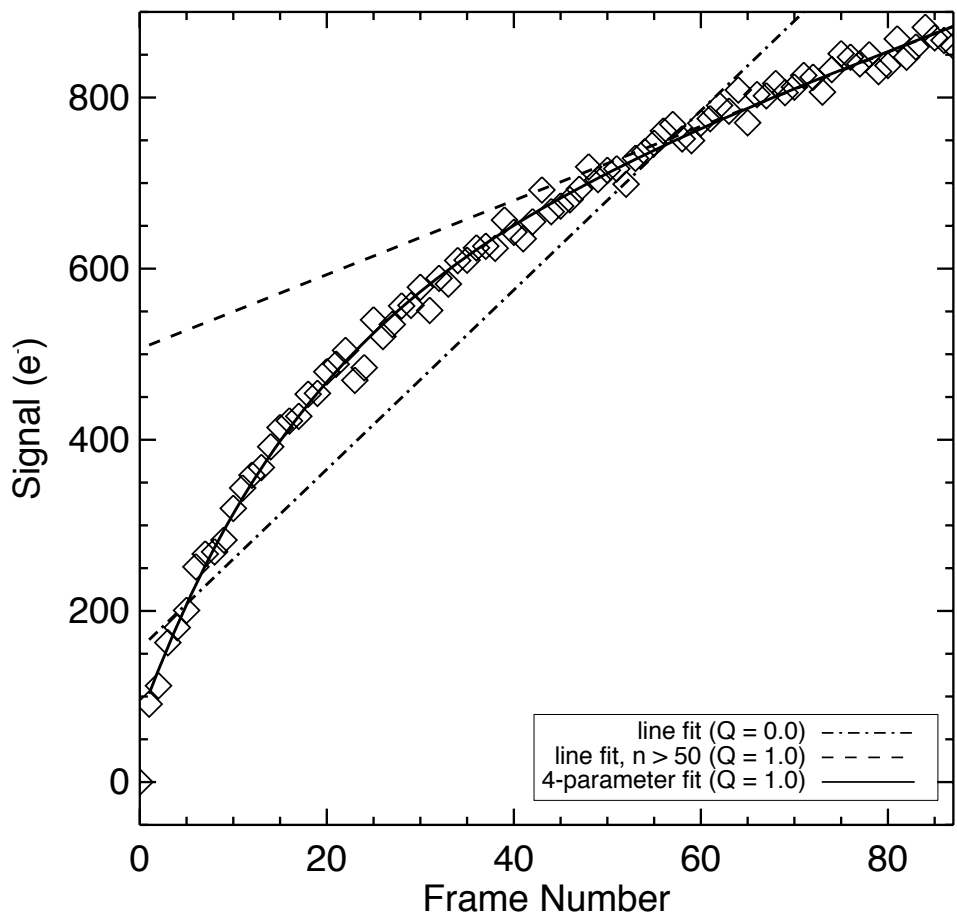}
\caption{The reset anomaly is a common nonlinear effect in the early
frames following pixel reset.  Here we show the 88 samples-up-the-ramp
for a pixel from engineering grade SCA \scas{016}.  The early samples
fall below the best fitting line drawn through later samples (dash).
If a linear fit is attempted through all the data points, the early
frames cause the fitted line (dash-dot) to over-estimate the dark
current.  The best fit for the entire data set (solid) indicates a four parameter 
equation that combines both exponential and linear terms.  
The goodness of fit is given by chi-square probability function, Q.} 
\label{fig_reset_fit} 
\end{center}
\end{figure}

It is not uncommon to observe a reset anomaly in MULTIACCUM sampled data
from JWST H2RGs (Figure~\ref{fig_reset_fit}).  The anomaly is
characterized by non-linearity in the early frames following pixel
reset.  Although the reset anomaly appears to be unrelated to response
linearity\footnote{For NIRSpec, we plan to confirm this by test of the
integrated DS.}, these early frames nonetheless fall below below a line
projected through the later, asymptotic portion of the ramp.
Fortunately, the reset anomaly is nearly noiseless for JWST SCAs that
have been tested so far, and it usually subtracts out during dark or sky
subtraction. Nevertheless, its potentially detrimental side effects must
be considered for the most accurate measurement of dark current.

Depending on the part, we have found that the fraction of affected
pixels can range from just a few percent to a significant fraction of
the SCA. Tests of the engineering grade $\lambda_{\rm co}=5~\mu$m
NIRSpec SCA \scas{016} revealed that over 15\% of pixels could not be
satisfactorily modeled by a straight line ($Q_{\rm line} < 0.1$). Here,
$Q$ is the integrated chi-square probability density giving the
probability that the fit's $\chi^2$ could have been obtained by chance
fluctuation within the error bars \citep[Equation~6.2.3]{pre92}. On the
other hand, the reset anomaly was barely noticeable in at least one
outstanding prototype SCA, \sca{015}. This detector is one of four JWST
SCAs in regular use at the University of Hawaii 2.2-m
telescope~\citep{hal04}.

The reset anomaly can introduce systematic errors into dark current
measurements if it is not correctly accounted for.  As illustrated in
Figure~\ref{fig_reset_fit}, if a 2-parameter line is fitted through all
points, the early frames cause the fitted line to over-estimate the
asymptotic slope, and thereby the dark current. 

One common solution is to discard the first few frames of each
integration.  Clearly, this is an inefficient use of time. Furthermore,
complete and unbiased removal of the reset anomaly is non-trivial. For
JWST SCAs, the reset anomaly has been observed to have time constants
ranging from seconds to hours before the pixels reach the asymptotic
portion of the ramp.  Moreover, different pixels in the same SCA have
different time constants.  Even by discarding the first few frames, it
is difficult to consistently identify the asymptotic portion of the
ramp, and a systematic bias tending to over-estimate the dark
current remains.

A solution that does not require discarding data is to extract the
asymptotic slope using a function that allows for the reset anomaly
early in the ramp.  Recent JWST testing has demonstrated that MULTIACCUM
sampled data from pixels showing the reset anomaly can be well-modeled
by a 4-parameter function that includes linear and exponential
components.  We speculate that the exponential term may be related to RC
charging effects in the ROIC/detector components of the hybrid. The
equation is of the form, 
\begin{equation}
s_{x,y}\left( t \right) = a_{x,y} + b_{x,y} t + c_{x,y} \exp\left(
d_{x,y}t \right), \label{eq_4_para}
\end{equation} 
where $s_{x,y}$ is the integrating signal, $t$ is time, and $a_{x,y}$,
$b_{x,y}$, $c_{x,y}$, and $d_{x,y}$ are the four fitting parameters. The
parameters $c_{x,y}$ and $d_{x,y}$  are negative quantities.
\citet{bac04} used the same equation for modeling the dark current of
pixels in a $\lambda_{\rm co}\! =\! 9.1~\mu{\rm m}$ detector array made
by Teledyne when they were known as Rockwell Scientific. Of the
non-linear pixels ($Q_{\rm line} < 0.1$), more than 70\% are well fitted
by the 4-parameter model ($Q_{\rm 4-param} > 0.1$). Of the remaining
non-linear pixels, many were hot pixels or were corrupted by RTN (see
Section~\ref{sec-rtn}).

Figure~\ref{fig_reset_fit} shows a direct comparison of all three
fitting methods.  The data are taken from a single pixel in a dark
integration.  A linear fit of the entire ramp clearly overestimates the
dark current.  The linear fit of the asymptotic portion of the ramp and
the 4-parameter fit provide much better results.  Although both of these
methods are comparable in their quality of fit, the 4-parameter fit does
not require any data to be discarded.  Furthermore, the asymptotic
portion of the ramp does not have to be identified for each pixel in the
array.

\subsection{Noiseless Calibration of the Reset Anomaly}
\label{ra_sub}

\begin{figure} 
\begin{center}
\includegraphics[width=5in]{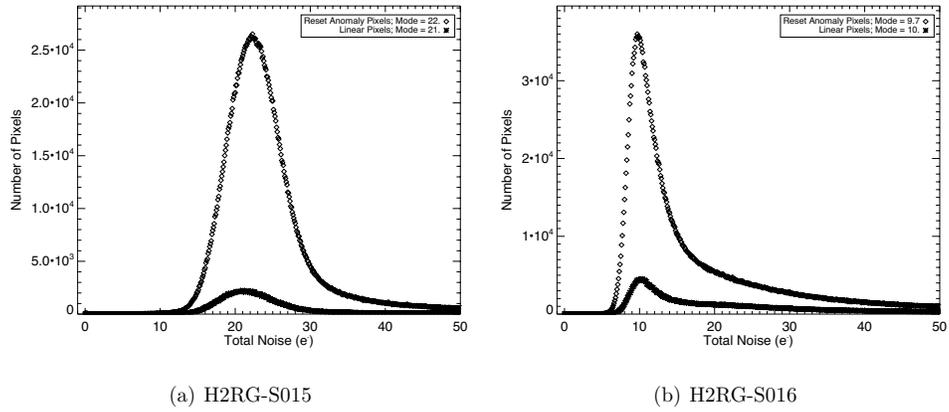}
\caption{The reset anomaly is nearly noise-less. Here we compare the
measured total noise for pixels having a significant reset anomaly to a
population of pixels that do not have the reset anomaly drawn from the
same SCA. Apart from normalization, the properties of the two
distributions do not differ significantly.}\label{fig-resetcmp}
\end{center}
\end{figure}

NIRSpec testing has shown that the reset anomaly is highly repeatable
for a given pixel.  A direct comparison of populations of pixels that
both are and are not affected by the reset anomaly indicates that the
reset anomaly contributes almost no additional noise
(Figure~\ref{fig-resetcmp}). Although the dark current properties of
these engineering grade SCAs are unacceptable for NIRSpec, the noise
properties of the two populations are essentially identical. 

We cross-checked these conclusions against science grade SCA~\sca{006}.
Although the available data sets do not allow us to make the same
statistical comparison that we make above for more recent parts, we have
compared the measured total noise using 88~samples taken at the
beginning of MULTI-145$\times$1 sampled integrations to 88~samples taken
at the very end. In this case, we find that using the first 88~frames
degrades the total noise by only a few percent compared to using the
last 88~frames. We used 88~frames as the basis of this comparison
because the NIRSpec baseline MULTI-22$\times$4 readout mode allows
88~frames per 1008 seconds integration.

\begin{figure}
\begin{center} 
\includegraphics*[angle=0,width=5in]{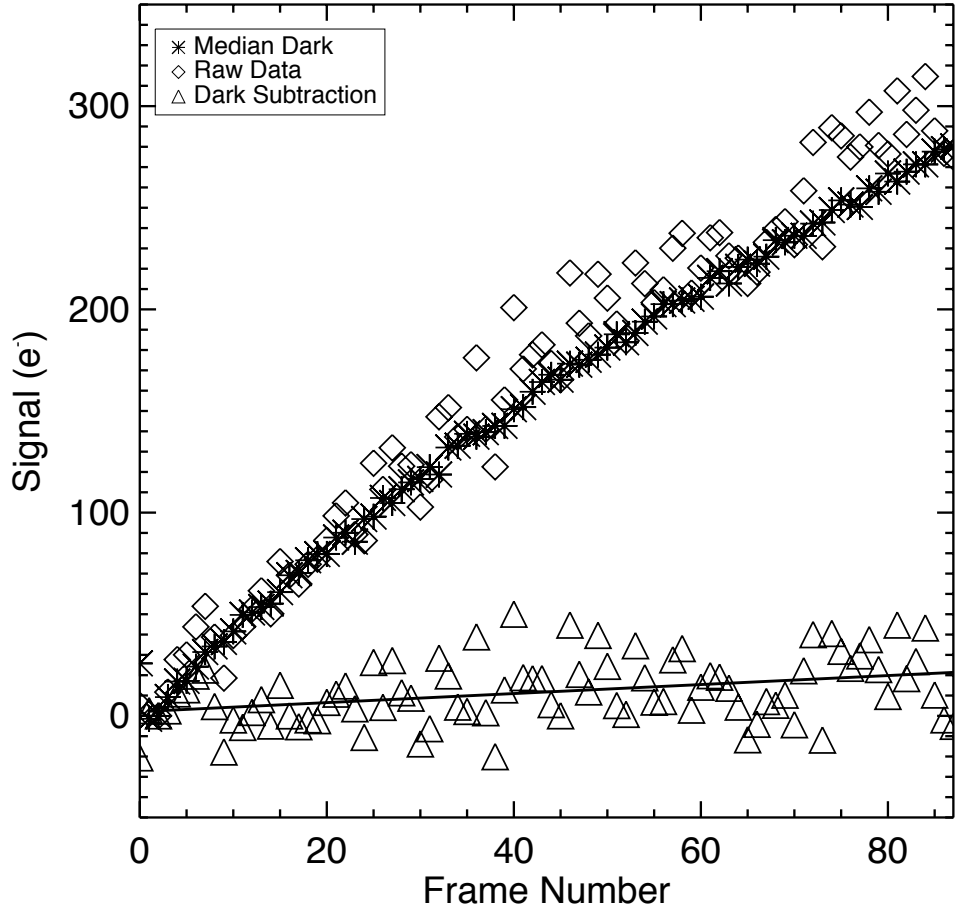}
\caption{The reset anomaly is a nearly noiseless instrument signature
that can be removed by subtracting a matching median dark cube (stars)
from an individual science integration (diamond). Here we show the 
88 samples-up-the-ramp for a pixel from engineering grade 
SCA \scas{016}. The data are shown before (diamond) and 
after (triangle) matching dark subtraction.}
\label{fig_darksub}
\end{center} 
\end{figure}

The reset anomaly calibrates out during matching dark or sky
subtraction. Figure~\ref{fig_darksub} shows the subtraction of a median
dark integration from an individual dark integration.  The subtraction
is performed using a matching MULTI-88$\times$1 median dark cube, which
was created from a median combination of 50 individual dark
integrations, pixel-by-pixel, within the 2048$\times$2048$\times$88
pixel cube.  The subtracted images have offsets and residual slopes,
which are the equivalent to $a_{x,y}$ and $b_{x,y}$, respectively, in
Equation~\ref{eqn-linemodel}.  The distribution of offsets is centered
at zero, which indicates that the reset anomaly has an identical shape
from one integration to the next.  The scatter in the offset, $a_{x,y}$,
is completely dominated by $kTC$ noise associated with resetting the
pixel at the beginning of the integration. In Section~\ref{ra_dark}, we
show the small residual slope is consistent with shot noise on
integrating dark current as predicted by Equation~\ref{eqn-multi-noise}
with $f\!=\!i_{\rm dark}$. 

\subsection{Unbiased Dark Current Measurements}
\label{ra_dark}

We tested the success of the 4-parameter model for measuring dark
current using real data from NIRSpec H2RGs.  In particular, we (1)
tested whether the dark current inferred from the 4-parameter fit could
account for the observed noise of the test SCAs and (2) compared the
success of the 4-parameter fit to the more traditional methods discussed
above.  These tests included a statistical analysis of the noise
properties of pixels in engineering grade NIRSpec SCAs \scas{015} and
\scas{016}.  We also performed less extensive spot checks on engineering
grade NIRSpec SCA \scas{002}.

We expect the measured total noise to be about equal to the noise
predicted by Equation~\ref{eqn-multi-noise}. The observed noise per
pixel is given by the standard deviation in the pixel's integrated
signal over many integrations.  We analyzed 50 individual integrations
taken in the DCL, as described in Section~\ref{dcl}.  To remove the
instrumental signature of the reset anomaly, we subtracted a median dark
integration from each individual integration.  As described in Section
\ref{ra_sub}, the reset anomaly is highly repeatable. A nearly noiseless
subtraction was obtained, as illustrated in Figure~\ref{fig_darksub}. 
The subtraction for each pixel generally results in a small residual
slope, $b_{x,y}$, with an offset, $a_{x,y}$.

To calculate the noise for each pixel ($x$,$y$), we fitted a 2-parameter
line to the residual slope in each of the 50 dark subtracted
integrations using Equation~\ref{eqn-linemodel}. The $a_{x,y}$ term,
which is completely dominated by $kTC$ noise, was discarded. The
$b_{x,y}$ term was used to calculate the integrated signal as follows, 
\begin{equation}
s_{x,y}=b_{x,y}t_{\rm int}. 
\end{equation}
The analysis produced 50 2-dimensional images of the residual signal. As
expected, the mean value of each pixel is zero $e^-$ to well within the
uncertainties. The noise of each pixel was computed as follows,
\begin{equation}
\sigma_{\rm total}\left[ x,y \right]= \left( \frac{1}{n-1}\sum_{i=1}^n
\left( s_i\left[ x,y \right] - \left< s\left[ x,y \right]\right>
\right)^2 \right)^{1/2} {\rm where~n=}50.
\label{eqn-test-noise} 
\end{equation}

Ideally, we expect the measured noise (Equation~\ref{eqn-test-noise}) to
equal the modeled total noise (Equation~\ref{eqn-multi-noise}).  In
other words, the ratio of measured to model noise values should be 1.0.
In Equation~\ref{eqn-multi-noise}, the variable $f$ is the dark current
of each pixel measured using the 4-parameter fit.  The read noise per
frame, $\sigma_{\rm read}$, is approximated using the spatial averaging
technique.  In spatial averaging, two correlated double sampling (CDS)
integrations, INT0 and INT1, are used to infer the average noise.  Each
CDS integration is represented by a data cube.  The first two dimensions
are the ($x$,$y$) pixel position, and the 3rd dimension gives the sample
number which can have the value 0 or 1.  $\sigma_{\rm read}$ was
calculated as follows,
\begin{equation} 
\sigma_{\rm read}^2  = \frac{1}{2} {\rm stdev} \left( \left( {\rm
INT1}\left[*,*,1\right]  - {\rm INT1}\left[*,*,0 \right] \right) -
\left( {\rm INT0} \left[*,*,1 \right] - {\rm INT0} \left[ *,*,0 \right]
\right) \right). \label{eq_sigma_CDS} 
\end{equation} 
Because statistical outliers can corrupt spatial averaging noise
measurements, iterative sigma clipping with a 3$\sigma$ threshold was
used to reject outliers.

\begin{figure}
\begin{center} 
\label{fig-noise-histo}
\includegraphics*[width=5in]{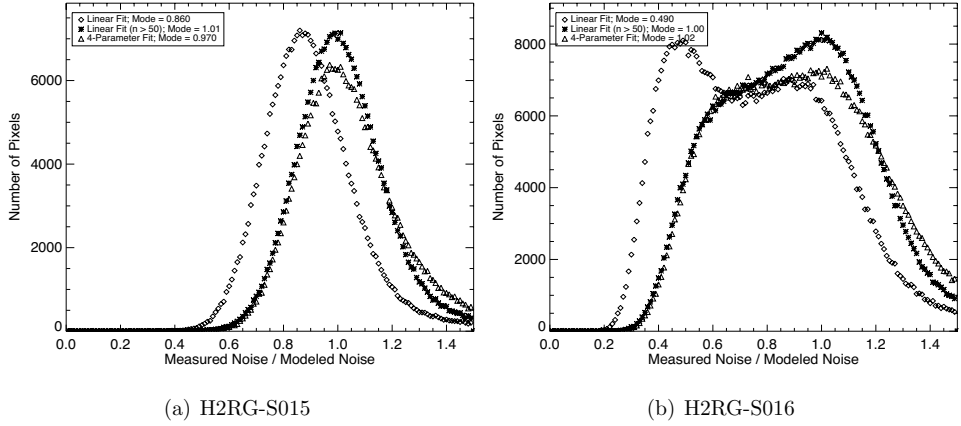}
\caption{These plots provide histograms of the ratio of the measured
noise to modeled noise for pixels in an SCA that can be characterized by
the reset anomaly.  The x-axis represents a pixel's average ratio taken
from 50 individual integrations.  The y-axis is the frequency of the given
ratio. The measured noise is calculated from
Equation~\ref{eqn-test-noise}, where $\sigma_{\rm total}$ is the
standard deviation in a pixel's signal over 50 individual integrations. 
The modeled noise is derived from Equation~\ref{eqn-multi-noise}, where
$f$ is the measured dark current in an individual data ramp.  The three
populations represented are the three different methods of measuring
dark current: a linear fit on the entire ramp, a linear fit on the
asymptotic portion of the ramp, and the 4-parameter fit. The latter two
provide a very good estimate of the dark current, while the linear fit
of the entire ramp tends to overestimate the linear slope.}
\end{center} 
\end{figure}

We analyzed the noise characteristics of pixels with the reset anomaly
in SCAs \scas{015} and \scas{016}. The dark current used in
Equation~\ref{eqn-multi-noise} was obtained from the 4-parameter fit.
For each pixel, the measured noise was compared to the mean predicted
noise.  The results are shown in Figure~\ref{fig-noise-histo}. The
success of the 4-parameter fit is highlighted by the agreement between
the measured and modeled noise values.  The ratio of the two noise terms
for SCAs \scas{015} and \scas{016} are 0.97 and 1.02, respectively.
These ratios are for the modes of the distributions.

For comparison purposes, the dark current was also measured using the
other fitting techniques described above: (1) linearly fitting the
entire ramp and (2) linearly fitting the asymptotic portion at the end
of the ramp. For consistency, the asymptotic portion of the ramp was
designated to be sample numbers greater than 50.  The results in
Figure~\ref{fig-noise-histo} indicate that a linear fit of the entire
ramp is a poor estimate of the dark current.  The measured and modeled
noise values do not agree within an acceptable uncertainty.  The linear
fit of the asymptotic portion at the end of the ramp does much better. 
The results are comparable to the 4-parameter fit. The ratio of the two
noise terms for SCAs \scas{015} and \scas{016} are 1.01 and 1.00,
respectively.  While this method provides adequate results, it requires
data to be discarded and does not provide consistent results due to
varying time constants.

While we are encouraged by the excellent agreement between measured and
modeled noise for these SCAs, this agreement depends in part on the
conversion gain, $g$. As explained in Section~\ref{sec:g}, conversion
gain was measured using the photon transfer method \citep{jan87}, and
for consistency in this argument we used the mode of the distribution of
$g$ values for each SCA.  Ideally, $g$ would be individually measured
for each pixel, and an IPC correction would be applied. Doing this
accurately requires larger data sets than are available for these
engineering grade parts, and better knowledge of the IPC than is
available at the present time. We therefore plan to revisit the
agreement between measured and modeled noise as more complete data sets,
including good measurements of IPC, become available for NIRSpec's
flight and flight spare SCAs in late 2007 and 2008.

\subsection{Note on Obtaining Convergence in 4-Parameter Fitting}
\label{sec-convergence}

We used the IDL procedure CURVEFIT for 4-parameter fitting.
Unfortunately, we find that it is often necessary to have good
first-estimates of the 4-parameters in advance of fitting a pixel to
ensure convergence. For the statistical analysis that are reported here,
a small set of pixels was studied to determine reasonable starting
coefficients for all pixels in the data set. A fully automated approach
is clearly preferable, and we plan to explore this further in future
publications.

\section{Random Telegraph Noise}
\label{sec-rtn}

In this section, we show that large-amplitude RTN affects a small and
fixed population of pixels. This confirms a previous finding by C.
McMurtry (pers. com. 2004). We believe that small-amplitude RTN, close
to the noise floor of the SCA, can probably be tolerated so long as it
does not cause pixels to exceed their stringent total noise budgets. If
substantiated by future testing of NIRSpec flight SCAs, we plan to
monitor and track RTN using standard pixel operability maps.

RTN has been observed in several JWST H2RG SCA's, as well as in four
H1RGs at the University of Rochester~\citep{bac05}. RTN is characterized
by a digital-like toggle between two (or more) levels.  For this reason,
RTN has also been referred to as ``popcorn mesa noise'' \citep{rau04}
and ``burst noise'' \citep{bac05}. Because RTN has been observed in both
regular and reference pixels, the noise is thought to originate in the
ROIC.  One likely explanation points to single-charge defects in the
unit cell MOSFET, which is the first amplifier seen by a detector diode.

\begin{figure}
\begin{center} \includegraphics*[angle=0,width=5in]{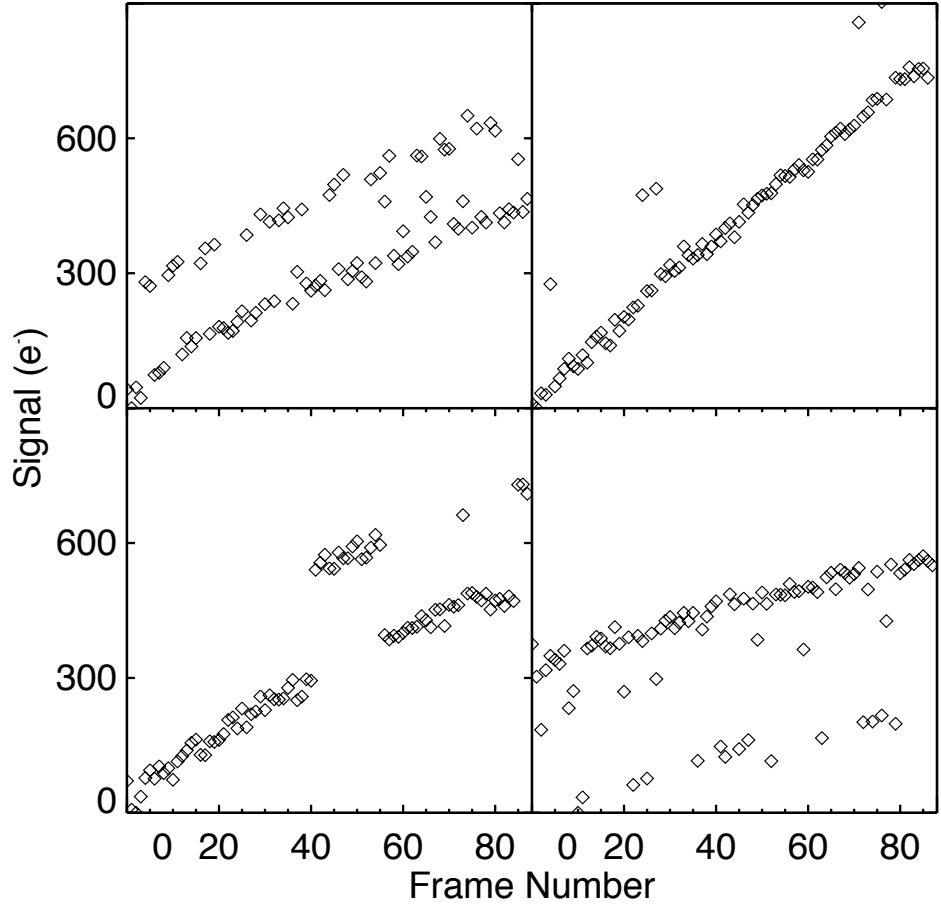} 
\caption{Random Telegraph Noise (RTN) is an artifact characterized by a
digital toggle between two (or more) signal levels.  This figure illustrates
the different patterns that RTN has been observed to exhibit.  While the
magnitude and frequency of the toggle varies between pixels, the noise
is consistent for a given pixel from integration to integration. RTN is
thought to arise from single-electron trapping effects in the ROIC.}
\label{fig_rtn} 
\end{center}
\end{figure}

Figure~\ref{fig_rtn} illustrates a few manifestations of RTN in JWST
H2RG pixels.  In each case, the data are distributed between two (or
more) distinct states.  The distribution characteristics of these
states, however, vary from pixel to pixel.  In particular, the states
can vary in size, and the frequency and magnitude of the scatter.

These variations make the detection of RTN difficult and time consuming.
We have developed a simple algorithm to detect RTN pixels in MULTIACCUM
sampled data. The algorithm consists of a two step process designed to
identify pixels that share the following two characteristics: (1)
unusually noisy sample ramps and (2) sharp rises and falls associated
with the digital toggle between the two states.

\begin{figure} 
\begin{center}
\includegraphics[width=3.9in]{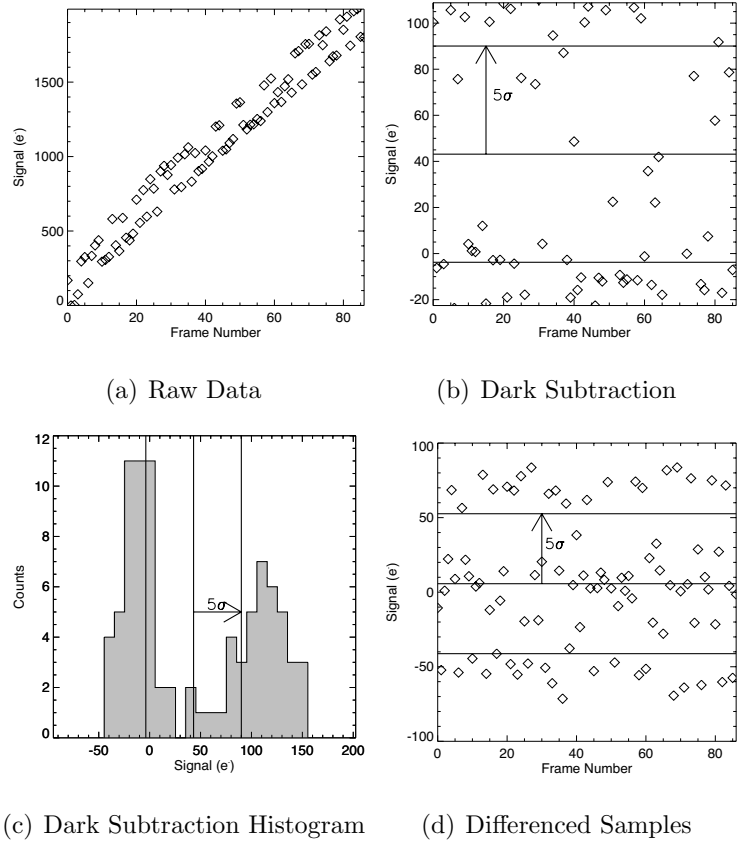} 
\caption{This figure illustrates the algorithm we have developed to
locate pixels that exhibit RTN, such as the one shown
in~\ref{fig_rtn_detector}a. The algorithm consists of a two step
process.  First, we identify noisy pixels, which we define to have
samples beyond  $\pm$5$\sigma_{\rm avg}$, where $\sigma_{\rm avg}$ is
the average scatter in the ramps.  To remove any offsets and correlated
noise effects, a median dark is subtracted from the individual
integration \ref{fig_rtn_detector}b.  For RTN pixels, two distinct
states are apparent by visual inspection, but can be more clearly
identified by the histogram in \ref{fig_rtn_detector}c.  To
differentiate between RTN and other noise effects, we then difference
successive data samples in order to identify the digital toggle
associated with the two (or more) states \ref{fig_rtn_detector}d. 
Again, a similar $\pm$5$\sigma_{\rm avg}$ threshold is used.  The
5$\sigma_{\rm avg}$ threshold was chosen in order to best isolate RTN
from other noise effects.  Therefore, this algorithm provides a lower
limit on the number of RTN pixels.}
\label{fig_rtn_detector} 
\end{center}
\end{figure}

The first step identifies noisy ramps.  Consider a typical pixel with
RTN (e.g. Figure~\ref{fig_rtn_detector}a).  To remove any offsets and
correlated noise effects, a median dark integration is subtracted from
the individual integration (Figure~\ref{fig_rtn_detector}b). The noise
in this ramp is revealed by the large degree of scatter.  Two distinct
readout states are revealed.  While these two states are apparent in
Figure~\ref{fig_rtn_detector}b by inspection, they are more clearly
illustrated by the histogram in Figure~\ref{fig_rtn_detector}c. The
scatter in these pixels tends to be larger than the average scatter,
$\sigma_{\rm avg}$. We flag all pixel ramps with a sample scatter beyond
$\pm$5$\sigma_{\rm avg}$ as potential RTN pixels.  Although this high
threshold has the advantage that it results in few false detections, it
also means that we miss smaller amplitude RTN pixels.

This first step, however, cannot distinguish between RTN pixels and
pixels that are naturally noisy.  The algorithm tends to return false
detections due to ``hot'' pixels that do not necessarily exhibit the two
(or more) distinct states that are associated with RTN.  These pixels
have a high degree of scatter because they typically have high dark
current and poor median dark subtraction.  For future detector
operation, we expect to have pixel masks which will allow us to identify
and avoid these ``hot'' pixels.  At the time of this analysis, however,
we implemented a second step to isolate RTN pixels.

This second step identifies pixel ramps that exhibit sharp, distinct
rises and falls.  This characteristic is typical of RTN, which is
identified by the toggling between two (or more) levels.  In comparison,
the noise in ``hot'' pixels is due to large dark current and does not
tend to toggle up and down.  Instead, the charge increases steadily,
just as it does in well-behaved pixels.  The only difference is that the
increase tends to be larger.  Differencing successive data points
provides an easy analysis of the pixel behavior.  The toggle in an RTN
pixel will produce a differential plot similar to the one shown in
Figure~\ref{fig_rtn_detector}d.  Again, the pixel differentials will
have an average scatter, $\sigma_{\rm avg}$.  Of these pixels flagged in
step one, all ramp differentials with scatter beyond $\pm$5$\sigma_{\rm
avg}$ are flagged as RTN pixels.

The success of this algorithm is highlighted by its false detection rate
of less than 1\%.  Nonetheless, we note that the algorithm's success is
limited by the chosen threshold.  For the present purpose of studying
RTN characteristics, we choose a $\pm$5$\sigma_{\rm avg}$ threshold to
best isolate pixels with RTN from pixels that may be affected by other
noise sources.  Therefore, our sample of RTN pixels represents a lower
limit on the actual number of RTN pixels within the array.  A ramp could
potentially have two states confined within the 5$\sigma_{\rm avg}$
threshold and would thereby go undetected. Setting the threshold lower
would increase the number of detections but it would also increase the
chance of a false detection due to the other sources of scatter.  A
possible solution utilizes multiple-Gaussian fitting to identify the two
unique populations apparent in
Figure~\ref{fig_rtn_detector}c~\citep{bac05}. 

\begin{figure} 
\begin{center}
\includegraphics*[width=5in]
{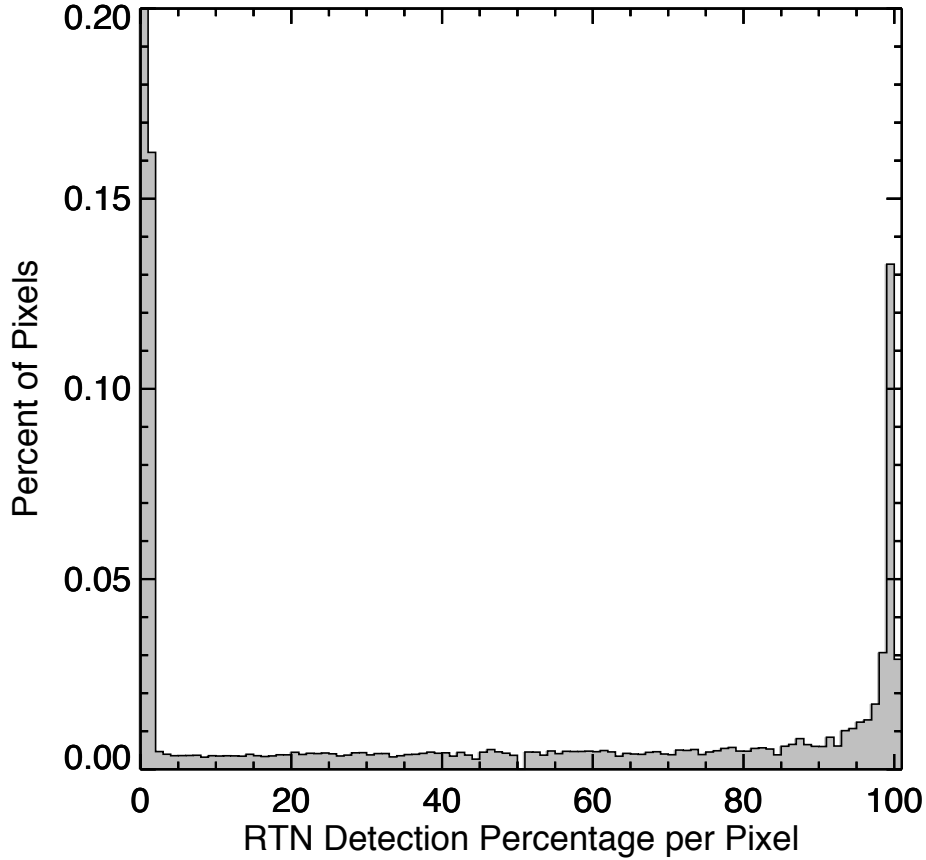} 
\caption{This histogram illustrates that RTN is largely confined to a
small and fixed subset of pixels, making it a feature that can be
tracked using operable pixel masks.  The peak at bin 0, which extends to
nearly 100\%, indicates that the vast majority of pixels have no
detectable RTN in any integration.  The peak at bin 99 indicates that of
pixels having detectable RTN in one integration, a majority have
detectable RTN in almost every other integration. The noticeable peak at
bin 1 and the drop off at bin 100, are due to the fluctuation in the
magnitude of the RTN scatter above and below the set thresholds. The
peak at bin 1 can also be partially attributed to the algorithm's $<$1\%
false detection rate.}
\label{fig_rtn_repeatability} 
\end{center}
\end{figure}

Using our 2-pass algorithm, we have observed large-amplitude RTN to
occur in a fixed, small subset of pixels.  For SCA \scas{16}, 99
integrations were tested. Figure~\ref{fig_rtn_repeatability} shows a
histogram which illustrates the repeatability of RTN detections per
pixel from integration to integration.  A vast majority of pixels have
zero detectable RTN features at the $\pm$5$\sigma_{\rm avg}$ threshold
in any of the 99 integrations sampled, as indicated by the peak at bin
0, which reaches beyond the extent of the plot to just under 100\%. 
Less than 1\% of pixels exhibited RTN characteristics at the
$\pm$5$\sigma_{\rm avg}$ threshold.  For a majority of those that did,
RTN was subsequently detected in that pixel for 99\% of integrations, as
indicated by the peak at bin 99.  The noticeable rise in bin 1 and fall
off in bin 100 is a result of the statistical nature of the magnitude of
the scatter. These features can also be partly attributed to the
algorithm's $<1$\% false detection rate.

For the engineering-grade JWST SCAs that have been studied to-date,
these results for \scas{16} are typical, and only a small percentage of
pixels appear to show large-amplitude RTN at T=37~K. Using a more
sensitive detection algorithm, \citet{bac05} found that 11\% of the
pixels in the SCA that they tested manifested RTN at T=37~K, and
moreover that there were significant temperature dependencies. These
included the size of the largest transition decreasing with increasing
temperature \citep{bac05}. The difference in the percentage of RTN
pixels reflects differences in detection algorithms, and possibly
device-to-device variation. 

As science and flight grade SCAs become available for JWST, we plan to
continue and extend these studies of RTN. One interesting conjecture is
that there may be a continuum of pixels affected by RTN (blending into
the read noise), and that the lower one sets the threshold, the more RTN
pixels one finds. Even if this conjecture were substantiated, however,
it is not clear to us that a pixel should be disqualified from use if it
meets all operability requirements while manifesting low-level RTN. At
some level, RTN becomes one of many components that contribute to the
overall noise of a pixel. Viewed in this light, RTN is a noise component
that has the advantage that it is easily identified, and can therefore
be fixed in future SCA designs.

The repeatability of large-amplitude RTN is good news.  The feature is
typically one of the noise components that can cause a pixel to fail to
meet operability requirements.  Locating and handling RTN pixels in real
time pipelined processing is costly and inefficient. Because
large-amplitude RTN is confined to a fixed, small subset of pixels, it
is a feature that can be tracked using a pixel operability mask. Because
tracking operable pixels is a standard part of calibration for flight
instruments, we expect large-amplitude RTN to have a negligible impact
on JWST calibration pipelines.

\section{Suggestions \& Plans for Future Work}

Additional study is needed to understand how repeatable small-amplitude
RTN is. Although we hypothesize that small-amplitude RTN is also a
property of a fixed population of pixels, it would be good to confirm
this by test. Doing this correctly requires a better RTN detection
algorithm than we have at the current time, and we plan to test this
hypothesis as better detection algorithms are developed.

Likewise, it would be helpful to know exactly where in the signal chain
RTN arises. We know that a significant fraction of the RTN, perhaps all
of it, originates in the ROIC. We know this because we see RTN in both
reference pixels, which are not connected to the HgCdTe detectors, and
regular pixels. Others have also used specialized readout software to
show that RTN originates in the ROIC \citep{bac04}. Simple physical
arguments suggest that the origin lies in the first MOSFET in the signal
chain, although it would clearly be better to experimentally pinpoint
the origin. Doing this could facilitate design improvements to eliminate
the RTN.

For similar reasons, it would be helpful to identify the physical
mechanism that is the underlying cause of the reset anomaly. As with
RTN, additional study would be helpful. One area that we plan to explore
more fully is whether the reset anomaly alters a pixel's response to
light. Although there has been no clear evidence of this in the JWST
program so far, it will be tested when we characterize the linearity
and photometric stability of the DS.

\section{Summary}

In this paper, we describe the JWST NIRSpec's baseline MULTIACCUM
readout mode, present a general noise model for NIR detector data
acquired using multiple non-destructive reads, and discuss recent
NIRSpec SCA test results. We believe that the noise model is applicable
to most astronomical NIR instruments. Our major findings and
recommendations are as follows.
\begin{enumerate} 
\item{The total noise in common NIR detector operating modes, including
CDS, MCDS (Fowler-N), and MULTIACCUM, can be modeled using
Equation~\ref{eqn-multi-noise} and the parameters listed in
Table~\ref{tab-model-params}. This noise model includes read noise, shot
noise on integrated charges, and covariance terms between multiple
non-destructive reads. If these covariance terms are neglected, and read
noise and shot noise are simply added in quadrature, we show that errors
of $\approx$9.5\% in the predicted noise for bright sources are
possible. The sense of the error is to under-predict noise when
covariance terms are neglected.}
\item{Many NIRSpec H2RG SCAs have shown a reset anomaly. This appears as
non-linearity in the early reads following reset. Although the reset
anomaly does not appear to be related to response linearity, we plan to
verify this by test for NIRSpec. If the reset anomaly is not correctly
accounted for during calibration, it can lead to systematic
over-estimation of the dark current. We show how the reset anomaly can
be noise-lessly calibrated out using matching darks, and how dark
current can be accurately measured in the presence of the reset anomaly
using 4-parameter fits.} 
\item{As has previously been reported, NIRSpec H2RGs are often affected
by RTN. Using new test data, we show that large-amplitude RTN is often a
property of only a small and fixed population of pixels. For flight
operations, we plan to monitor and track RTN using pixel operability
maps.}
\end{enumerate}

These conclusions, particularly with regard to the reset anomaly and
RTN, are largely based on testing engineering grade SCAs. This was done
because the required large data sets are only available from engineering
grade parts at this time. We therefore plan to confirm these findings
using better SCAs as they become available.

\acknowledgments

We thank Judy Pipher, Craig McMurtry, and Bill Forrest for their many
thoughtful comments and suggestions during the preparation of this
manuscript. This research was supported by NASA and ESA as part of the
James Webb Space Telescope Project. Ori Fox wishes to thank NASA's
Graduate Student Researcher Program for a grant to the University of
Virginia.


\clearpage
\begin{deluxetable}{lcl}
\tablewidth{0pt}
\tablecaption{Driving NIRSpec Detector Performance
Requirements\label{tab-detreq}}
\tablehead{\colhead{Parameter}&
\colhead{Requirement}& \colhead{Comment}} 
\startdata 
Total noise ($e^-$~rms)& 6& $t_{\rm int}=1008~{\rm s}$\\ 
&& multi-22$\times$4\\ Mean
dark current ($e^- {\rm s}^{-1} {\rm pixel}^{-1}$)& 0.010\\ 
DQE& 70\%&
$0.6\le\lambda < 1.0 \mu$m\\ 
& 80\%& $1\le\lambda < 5 \mu$m\\ 
Operating temperature (K)& 34-37\\ 
Pixel operability for science\tablenotemark{1}& $>$92\%
\enddata 
\tablenotetext{1}{Pixel operability for science includes stringent
thresholds on total noise and DQE. Pixels that fail to meet the
operability for science requirement are degraded, although they may
still be useful for target acquisition and other less sensitive
observations.}
\end{deluxetable}

\clearpage
\begin{deluxetable}{lccl} \tablewidth{0pt}
\tablecaption{Model Parameters for Common Readout
Modes\tablenotemark{1}\label{tab-model-params}} 
\tablehead{
\colhead{Readout Mode} & \colhead{$n$} & \colhead{$m$} &
\colhead{Comments}} 
\startdata
MULTI-22$\times$4 & 22 & 4 & JWST NIRSpec baseline\\
MULTI-6$\times$8 & 6 & 8 & JWST NIRCam baseline\\ 
CDS& 2& 1& Correlated double sampling\\ 
MCDS-8\tablenotemark{2}& 2& 8&  Also known as Fowler-8\\
MCDS-16& 2& 16& Fowler-16\\ 
MCDS-32& 2& 32& Fowler-32\\ 
\enddata
\tablenotetext{1}{For many astronomical detector arrays, the read noise
per frame is approximately $\sigma_{\rm read}\!\approx\!\sigma_{\rm
CDS}/\sqrt{2}$. This approximation is appropriate for short dark
integrations, for which shot noise on integrated dark current is
negligible compared to read noise.} 
\tablenotetext{2}{For MCDS readout modes, $t_g\!=\!t_{\rm int}$.}
\end{deluxetable}

\clearpage
\begin{deluxetable}{lccccccccc}
\rotate
\tablewidth{0pt}
\tablecaption{Summary of JWST NIR
SCAs\tablenotemark{a}\label{tab-sca-summary}}
\tablehead{
&
&
\colhead{$\sigma_{\rm CDS}$}&
\colhead{$\sigma_{\rm total}$}&
\colhead{$i_{\rm dark}$}&
\multicolumn{2}{c}{QE}&
&
\\
\colhead{Serial Number}&
\colhead{Grade}& 
\colhead{($e^-$~rms)}& 
\colhead{($e^-$~rms)}& 
\colhead{($e^-~{\rm s^{-1}~pixel^{-1}})$}&
\colhead{1.25~$\mu$m}&
\colhead{2.2~$\mu$m}&
\colhead{Crosstalk}&
\colhead{Persistence}
}
\startdata
\sca{006}&
sci&
12.5\tablenotemark{c}&
6\tablenotemark{b,c}&
.004\tablenotemark{c}&
\nodata&
\nodata&
\nodata&
\nodata\\
\sca{015}& 
sci& 
\nodata&
5.88\tablenotemark{d}&
.006\tablenotemark{c}& 
95\%\tablenotemark{c} &
95\%\tablenotemark{c} &
1.56\%\tablenotemark{d}&
0.1\%\tablenotemark{d}\\
\scas{015}&
eng&
12.3\tablenotemark{e}&
16.5\tablenotemark{e}&
0.28\tablenotemark{e}&
\nodata&
\nodata&
\nodata&
\\
\scas{016}&
eng&
14.5\tablenotemark{e}&
8.6\tablenotemark{e}&
0.004\tablenotemark{e}&
\nodata&
\nodata&
\nodata
\enddata
\tablenotetext{a}{All tests were performed at T=37~K. The detectors
were biased to meet the NIRSpec well-depth requirement of $6\times
10^4~e^-$.}
\tablenotetext{b}{MCDS-16 sampling (Fowler-16) was used for this early
measurement. For all other $\sigma_{\rm total}$ measurements, which were
made later, NIRSpec-baseline MULTI-22$\times$4 sampling was used.}
\tablenotetext{c}{\citet{rau07}}
\tablenotetext{d}{\citet{fig04}}
\tablenotetext{e}{NASA GSFC Detector Characterization Laboratory
measurement.}
\end{deluxetable}


\clearpage 

\clearpage

\clearpage

\clearpage

\clearpage

\clearpage 

\clearpage 

\clearpage 

\clearpage

\clearpage

\end{document}